\documentclass[%
 aip,
rsi,%
 amsmath,amssymb,
reprint,
]{revtex4-1}

\usepackage{graphicx}
\usepackage{dcolumn}
\usepackage{bm}
\usepackage{tabularx}
\usepackage{setspace}
\usepackage{graphicx}
\usepackage[caption=false]{subfig}
 \usepackage{color}

\graphicspath{{./PAPER_FIGURES/}}

\begin{document}

\preprint{AIP/123-QED}

\title{A compact actively damped vibration isolation platform for experiments in ultra-high vacuum}

\author{{\'A}lvaro Fern{\'a}ndez Galiana}
\email{alvarofg@mit.edu}
\affiliation{LIGO, Massachusetts Institute of Technology, Cambridge, MA 02139, USA}
\affiliation{Mechanical Engineering Department, MIT, Cambridge, MA 02139, USA}

\author{Lee McCuller}
\affiliation{LIGO, Massachusetts Institute of Technology, Cambridge, MA 02139, USA}

\author{Jeff Kissel}
\affiliation{LIGO Hanford Observatory, Richland, WA 99352, USA}

\author{Lisa Barsotti}
\affiliation{LIGO, Massachusetts Institute of Technology, Cambridge, MA 02139, USA}

\author{John Miller}
\affiliation{LIGO, Massachusetts Institute of Technology, Cambridge, MA 02139, USA}

\author{Maggie Tse}
\affiliation{LIGO, Massachusetts Institute of Technology, Cambridge, MA 02139, USA}
\affiliation{Physics Department, MIT, Cambridge, MA 02139, USA}

\author{Matthew Evans}
\affiliation{LIGO, Massachusetts Institute of Technology, Cambridge, MA 02139, USA}
\affiliation{Physics Department, MIT, Cambridge, MA 02139, USA}

\author{Stuart M. Aston}
\affiliation{LIGO Livingston Observatory, Livingston, LA 70754, USA}

\author{TJ Shaffer}
\affiliation{LIGO Hanford Observatory, Richland, WA 99352, USA}

\author{Arnaud Pele}
\affiliation{LIGO Livingston Observatory, Livingston, LA 70754, USA}

\author{Janeen H. Romie}
\affiliation{LIGO Livingston Observatory, Livingston, LA 70754, USA}

\author{Betsy Weaver}
\affiliation{LIGO Hanford Observatory, Richland, WA 99352, USA}

\author{Richard Abbott}
\affiliation{LIGO, California Institute of Technology, Pasadena, CA 91125, USA}

\author{Peter Fritschel}
\affiliation{LIGO, Massachusetts Institute of Technology, Cambridge, MA 02139, USA}

\author{Nergis Mavalvala}
\affiliation{LIGO, Massachusetts Institute of Technology, Cambridge, MA 02139, USA}
\affiliation{Physics Department, MIT, Cambridge, MA 02139, USA}

\author{Fabrice Matichard}
\email{fabrice@ligo.mit.edu}
\affiliation{LIGO, Massachusetts Institute of Technology, Cambridge, MA 02139, USA}
\affiliation{LIGO, California Institute of Technology, Pasadena, CA 91125, USA}

\date{\today}

\begin{abstract}

We present a tabletop six-axis vibration isolation system, compatible with Ultra-High Vacuum (UHV), which is actively damped and provides 25 dB of isolation at 10 Hz and 65 dB at 100 Hz. While this isolation platform has been primarily designed to support optics in the Laser Interferometer Gravitational-Wave Observatory (LIGO) detectors, it is suitable for a variety of applications. The system has been engineered to facilitate the construction and assembly process, while minimizing cost. The platform provides passive isolation for six degrees of freedom using a combination of vertical springs and horizontal pendula. It is instrumented with voice-coil actuators and optical shadow sensors to damp the resonances. All materials are compatible with stringent vacuum requirements. Thanks to its architecture, the system's footprint can be adapted to meet spatial requirements, while maximizing the dimensions of the optical table. Three units are currently operating for LIGO. We present the design of the system, controls principle, and experimental results.
\end{abstract}

\maketitle

\section{\label{sec:Introduction}Introduction}

\begin{figure} [b!]	
	\centering
	\includegraphics[width=0.96\columnwidth]{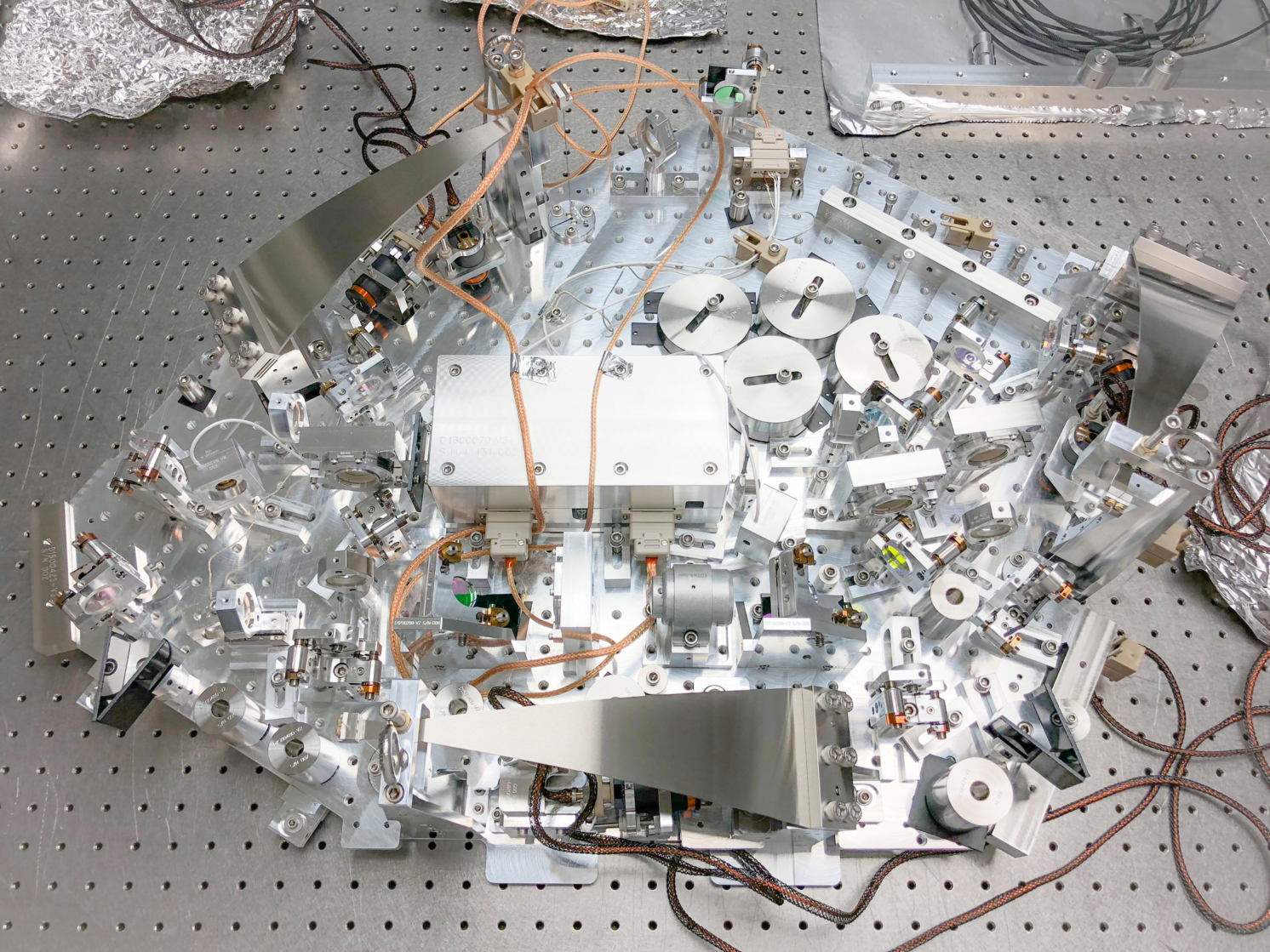}
	\caption{\label{fig:VOPO} \textbf{UHV COMPATIBILITY.} The isolation platform at the LIGO Livingston Observatory. All the materials fulfill stringent UHV requirements \cite{vacuum}.}
	\end{figure}

 Many precision measurement experiments must be performed in a seismically isolated environment \cite{multiaxis} and under vacuum \cite{UHV1,UHV2} to achieve their designed sensitivity. 
 We present an Ultra-High Vacuum (UHV) compatible tabletop 6-axis vibration isolation system, with integrated active damping control, that provides three orders of magnitude of isolation above 100 Hz. 
 This system has been designed to provide isolation to a new component of the Advanced LIGO (aLIGO) interferometers \cite{ligo}, a squeezed light source (the Vacuum Optical Parametric Oscillator) that aims to improve the sensitivity of the interferometer by reducing quantum shot noise \cite{shot,aLIGO}. Isolation is obtained by a combination of vertical springs and horizontal pendula, commonly used in Advanced LIGO \cite{matichard2015seismic}, while voice-coil actuators provide active damping. Unlike other Advanced LIGO vibration isolation suspensions \cite{HLTS,tiptilt}, the support structure is not based on a welded frame and takes advantage of independent blade posts, making its design simpler and more flexible. In particular, the dimensions of the optical table can be adapted to fit space constrains (see Fig. \ref{fig:VOPO}).
 
The paper is organized as follows: the mechanical design and both the horizontal and vertical isolation of the system are described in Sec. \ref{sec:MechDe}; active damping control topology is presented in Sec. \ref{sec:actdamp}; and in Sec. \ref{sec:perf} the results of experimental tests, performed on the first platform prototype unit assembled at the MIT-LIGO laboratory and on the two units subsequently installed at the LIGO observatories, are discussed.

\section{\label{sec:MechDe}Mechanical Design} 

\begin{figure}	[t!]
	\centering
	\includegraphics[width=0.96\columnwidth]{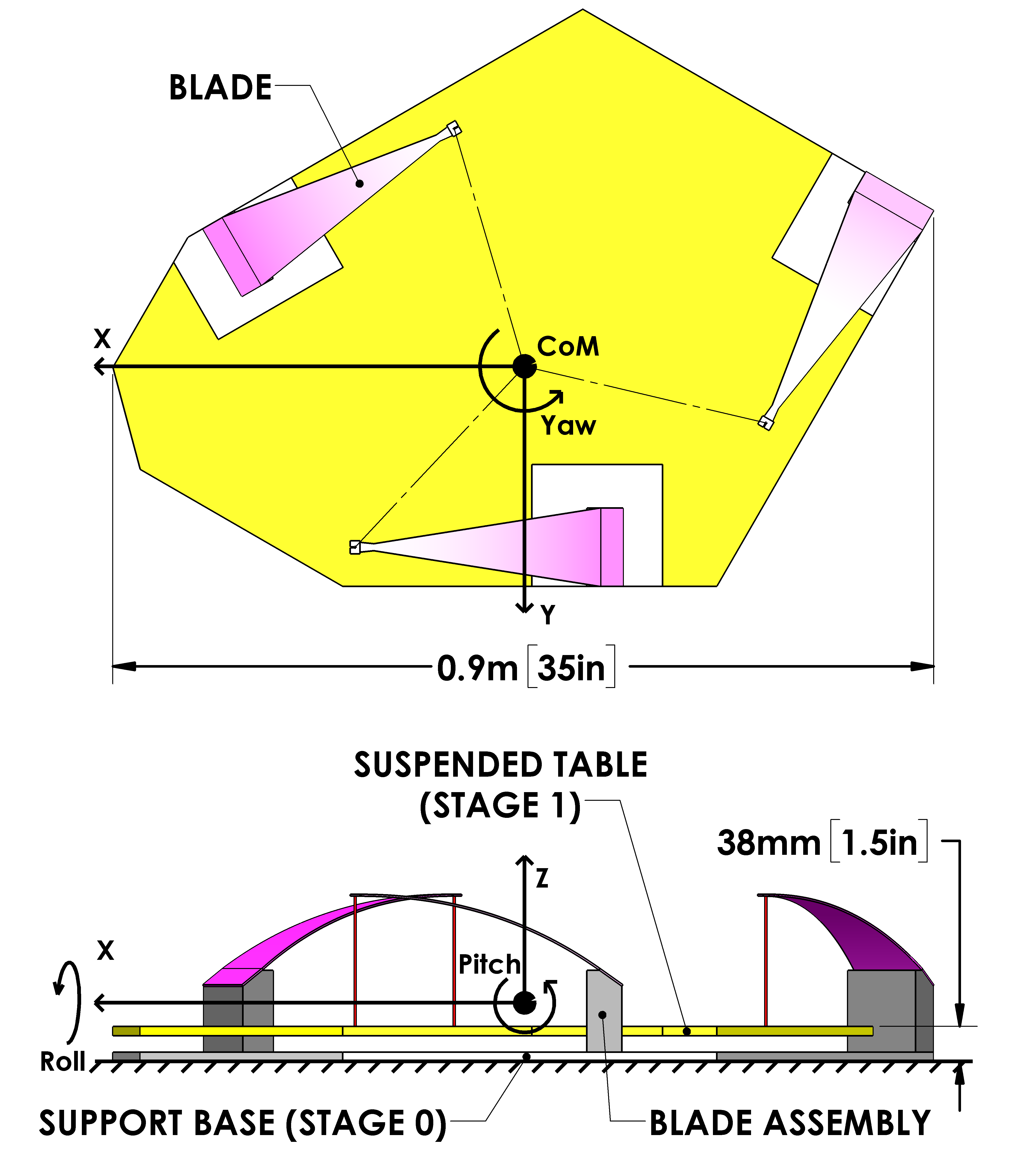}
	\caption{\label{fig:VOPO_Schematics}  \textbf{COMPACTNESS.} Schematics of the isolation platform. The optical table is suspended from three blade assemblies, 38 mm (1.5 in) above the reference plane. The footprint is designed to match the space in the LIGO chambers.}
	\end{figure}

The system provides seismic isolation and active damping control in all six degrees of freedom. The suspended stage (stage 1) is mechanically isolated from the support base (stage 0) (see Fig. \ref{fig:VOPO_Schematics}). 

\begin{figure} [t!]

\subfloat[\label{fig:Blade_Post_AB}]{\includegraphics[clip,width=0.96\columnwidth]{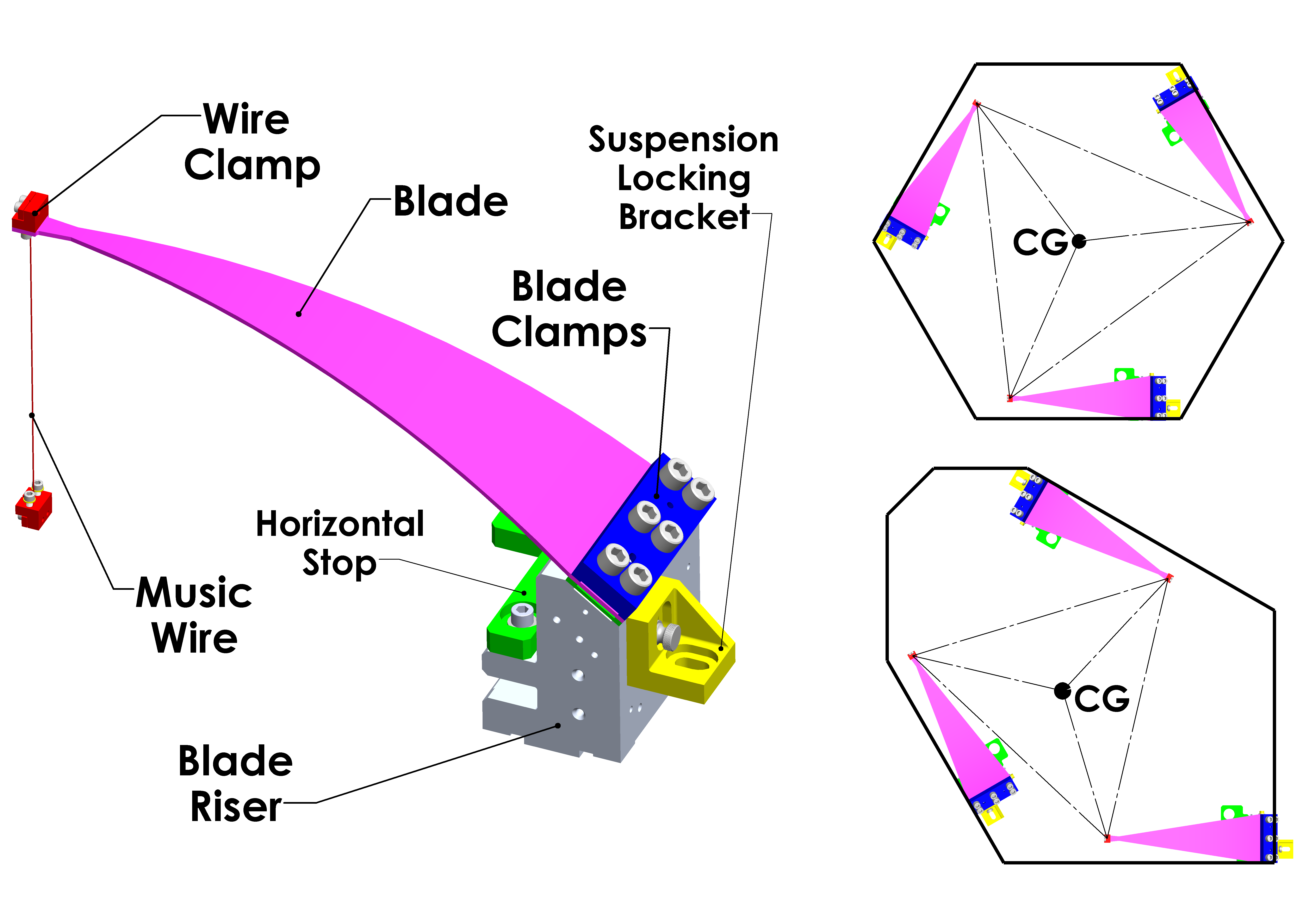}}%

\subfloat[\label{fig:VOPO_chamb}]{\includegraphics[clip,width=0.96\columnwidth]{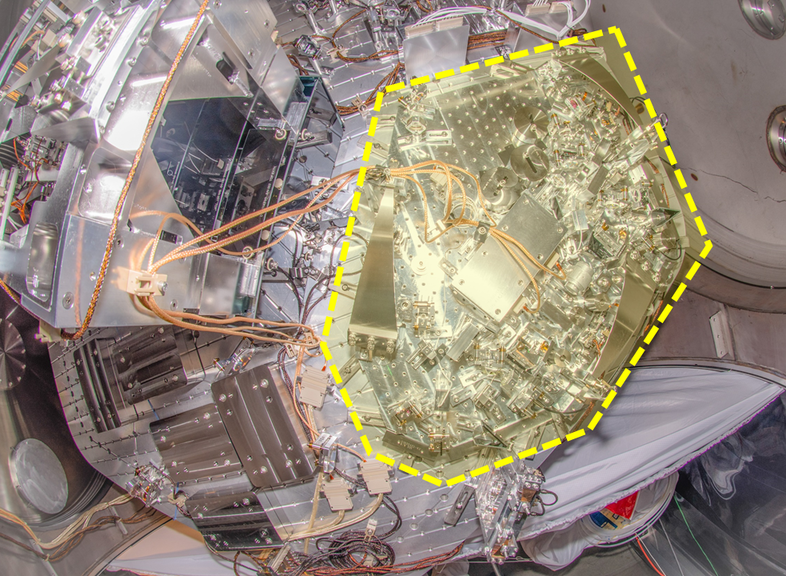}}%
  
  \caption{\label{fig:Blade_Post}\textbf{SHAPE ADAPTABILITY.} Blade assembly (a-left), including the riser clamped to the base,  the blade clamped at a specific launch angle, and the pendulum wire clamped to the tip of the blade. The blade assembly includes built-in hard stops that limit the range of motion of the optical table. Using independent blade assemblies to suspend the optical table makes the design adaptable to different requirements. As an illustration, two optical table shapes that can be implemented using this concept are presented (a-right). Our design was adapted to match the space available in the aLIGO chambers (b).}
\end{figure}

Stage 1 is suspended with three independent blade assemblies (see Fig. \ref{fig:Blade_Post}) bolted to stage 0. This approach, based on low-height posts, reduces the cost, weight, and dimensions compared to the welded support frames used in previous LIGO suspensions \cite{HLTS, tiptilt}. It maximizes the optical table dimensions with respect of the space available and makes the system adaptable to different optical table shapes (see Fig. \ref{fig:Blade_Post}). The number, position, and characteristics of the blade assemblies can easily be adapted to various table shapes, payloads, and performance requirements.

\begin{table} [b!]
\caption{\label{tab:Mechanical} Dimensions, mass and inertia parameters}
\centering
\begin{spacing}{1.5}
\begin{tabularx}{.94\columnwidth}{p{35mm}c}
 \hline \hline
 Parameter & Value \\ 
 \hline\hline
 Overall dimensions & 0.94 x 0.67 $\mathrm{m}$ \ [26.5 x 37 $\mathrm{in.}$] \\ 
 Total height  & \ 0.21 $\mathrm{m}$\ [8.1 $\mathrm{in.} $] \\
 Total mass & 45 $\mathrm{kg}$  \\ 
 \hline
Suspended stage: &   \\
 Table surface & 0.34 $\mathrm{m}^2$ \ [520 $\mathrm{in.}^2$] \\
 Table height &0.38 $\times 10^{-3} \ \mathrm{m}$ \ [1.5 $\mathrm{in.}$]  \\
 Mass ($m$) & 36 $\mathrm{kg}$  \\
 $z_{_\mathrm{CoM}}$ & 7.82 $\times 10^{-3} \ \mathrm{m}$ \\
 Inertia (Roll) & 0.705 $\mathrm{kg} \ \mathrm{m}^2$ \\ 
 Inertia (Pitch) & 1.313 $\mathrm{kg} \ \mathrm{m}^2$ \\
 Inertia (Yaw)  & 1.958 $\mathrm{kg} \ \mathrm{m}^2$\\
 \hline\hline
\end{tabularx}
\end{spacing}
\end{table}

\begin{figure}	[t!]
	\centering
	\includegraphics[width=0.96\columnwidth]{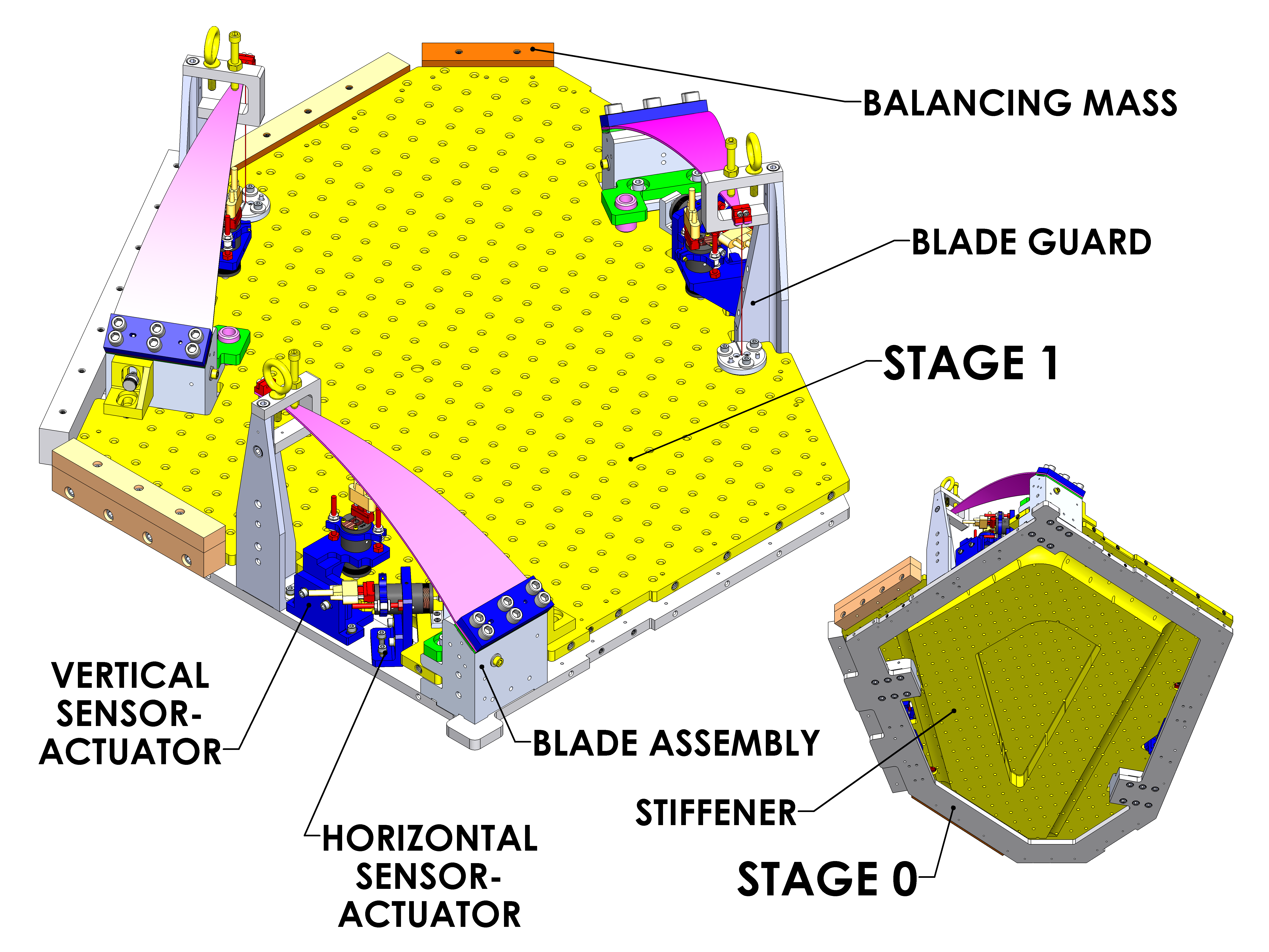}
	\caption{\label{fig:VOPO_CAD}\textbf{EFFECTIVE FOOTPRINT.} Most of the footprint corresponds to the optical table. The isolation platform is instrumented with six sensor-actuators positioned in vertical-horizontal pairs.  The blade guards provide safety during assembly. The optical table features a stiffener underneath to increase the stiffness to mass ratio.}
	\end{figure}

There are two types of resonances that are critical to the design of this type of seismic isolator: the suspension modes and the structural modes. The first are the rigid-body modes associated with each of the six degrees of freedom. The structural modes are the flexible modes of the sub-assemblies, such as the blade assemblies or blade guards (see Fig. \ref{fig:VOPO_CAD}). Since suspension modes have lower frequencies than structural modes, the isolation bandwidth can be defined as the range between the highest suspension mode frequency and the lowest structural mode frequency.

LIGO's requirements \cite{aLIGO} drive the choice of the highest suspension mode frequency ($\approx$2.5 Hz) and the lowest structural mode frequency ($\approx$200  Hz).  At 100 Hz, the suspension provides 65 dB of seismic isolation, and the suspension modes are actively damped to quality factors of about 20. The main parameters of the platform are specified in Table \ref{tab:Mechanical}.

\subsection{Ultra-high vacuum compatibility}
All the materials used in the isolation platform are approved for ultra-high vacuum applications and fulfill LIGO vacuum system specific requirements \cite{vacuum} (see Fig. \ref{fig:VOPO}). Before final assembly, all the components are precision cleaned and vacuum baked. In addition, a residual gas analysis is performed after baking to measure the outgasing rate and detect possible residual contamination.

\subsection{Suspended stage}
The suspended stage consists of an optical table ($\approx$0.34 m$^2$, $\approx$18 kg), optical components ($\approx$11 kg), and balancing masses ($\approx$7 kg) used to level the platform. The table shape is adapted to the available space in the LIGO vacuum chamber \cite{HAMISI1}, resulting in the heptagonal shape shown in Fig. \ref{fig:VOPO_Schematics}.

\begin{figure}	[b!]
	\centering
	\includegraphics[width=0.96\columnwidth]{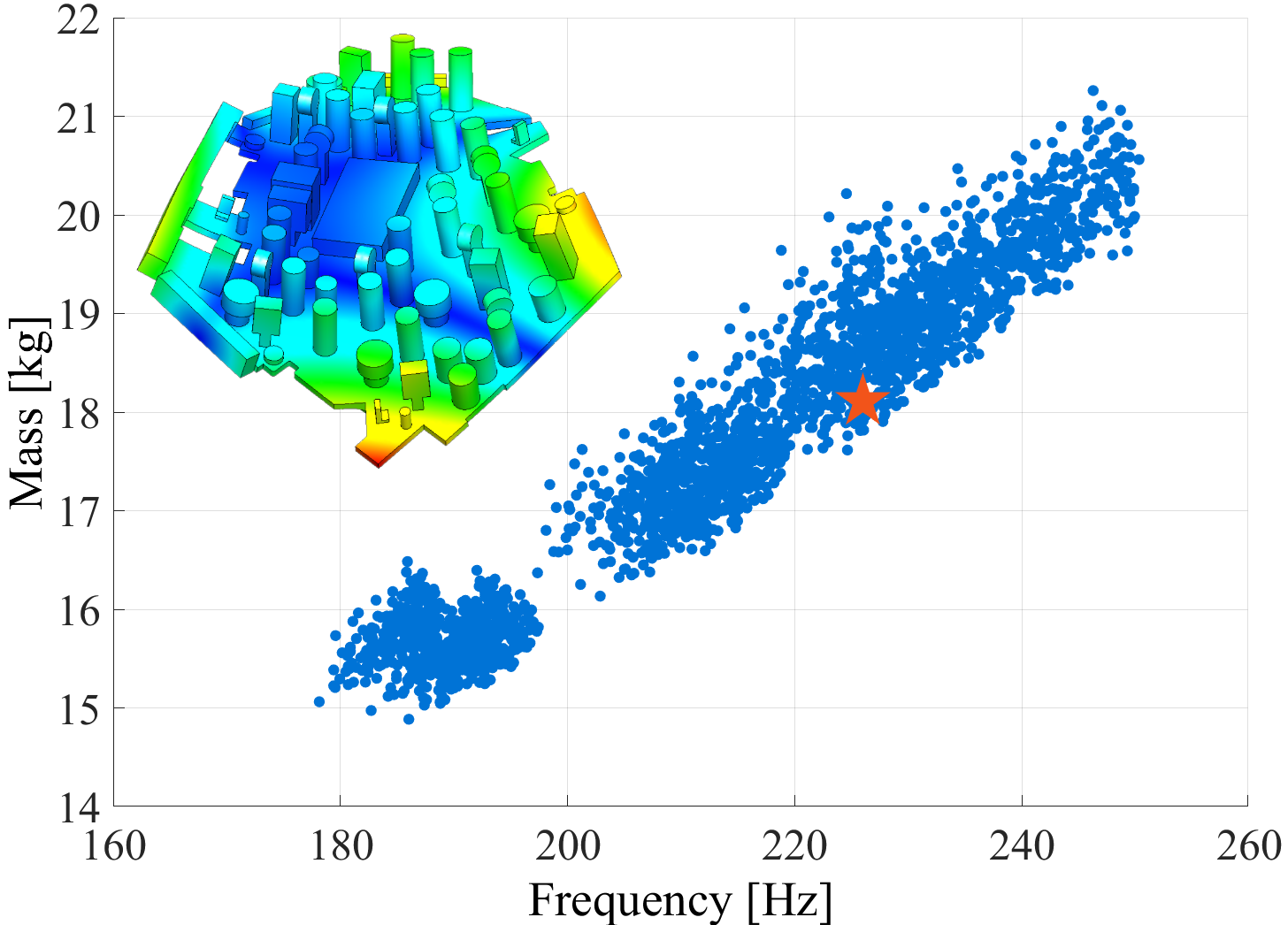}
	\caption{\label{fig:Bench_opti} \textbf{TABLE STIFFNESS.} Results of the table stiffness optimization. Using iterative finite element analysis on the optical table for different stiffener shapes and sizes. The selected dimension (indicated with a star) was chosen to have the highest stiffness to mass ratio while meeting the requirements on total mass and center of mass position.}
	\end{figure}

The first structural mode is a bending resonance of the suspended table. To raise its frequency, a stiffener is incorporated into the optical table (see Fig. \ref{fig:VOPO_CAD}). The shape of the stiffener has been designed to optimize the stiffness to mass ratio. This optimization is performed iteratively using Finite Element Analysis (FEA) of the loaded optical table. Fig. \ref{fig:Bench_opti} shows the results of this optimization, with the first resonance at $\approx$225 Hz.

\subsection{Horizontal isolation}

Horizontal isolation (X-Y-Yaw axis in Fig. \ref{fig:VOPO_Schematics}) is achieved by suspending stage 1 from three wires. The pendulum is a simple and effective way to provide passive horizontal isolation. The wires are attached using metal clamps designed to reduce friction loss. The horizontal modes frequencies can be approximated using Eq. (\ref{eqn:fxx}). 
	\begin{equation}
	f_{xx} \ = f_{yy} = \frac{1}{2 \pi}\sqrt{\frac{k_{xx}}{m}}
	\label{eqn:fxx}
	\end{equation}
	
	\noindent		
 where $k_{xx}$ and $m$ are, respectively, the horizontal stiffness and supported mass of each wire. The stiffness can be estimated using Eq. (\ref{eqn:kxx}).
	\begin{equation}
	k_{xx} \ = k_{yy} =\frac{m g}{l_w - 2 z_{_\mathrm{ZMP}}}
	\label{eqn:kxx}
	\end{equation}

	\noindent	
where $l_w$ is the length of the wires and $z_{_\mathrm{ZMP}}$ is the vertical distance from the wire tip to the Zero Moment Point (ZMP). This  corresponds to
the distance over which the wire bends near the clamps, and can be estimated by:  
	\begin{equation}
	z_{_\mathrm{ZMP}} = \sqrt{\frac{E_w I_w}{m g}} \ \tanh{\left(\frac{m g l_w}{2 E_w I_w}\right)}
	\label{eqn:ZMP}
	\end{equation}

	\noindent	
$E_w$ being the wire Young's modulus, and $I_w = \frac{\pi}{64}d_w^4$ its second moment of inertia.

The suspension wires are made of stainless steel; they are 146 mm long and 0.61 mm in diameter. These dimensions result in an horizontal suspension frequency of 1.27 Hz (Eq. (\ref{eqn:fxx})). 
The horizontal range of motion is limited by hard stops designed to prevent damage in the event of earthquakes. The maximum horizontal displacement allowed is $\pm$1 mm. Using the maximum displacement, we can compute the maximum normal ($\sigma_\mathrm{max}$) and shear ($\tau_\mathrm{max}$) stresses at the wire with Eq. (\ref{eqn:wire_stress_sigma}) and Eq. (\ref{eqn:wire_stress_tau}), respectively.
	\begin{equation}
    \sigma_\mathrm{max}^w = \frac{m g}{A_w} + \frac{m g z_{_\mathrm{ZMP}} d_w \delta}{2 I_w (l_w - 2 z_{_\mathrm{ZMP}})}
	\label{eqn:wire_stress_sigma}
	\end{equation}
	\begin{equation}
	\tau_\mathrm{max}^w =  \frac{m g \delta}{A_w (l_w - 2 z_{_\mathrm{ZMP}})}    
	\label{eqn:wire_stress_tau}
	\end{equation}

\noindent
where $\delta = 1$mm is the maximum allowed lateral displacement, $d_w$ is the diameter of the wire, and $A_w = \frac{\pi d_w^2}{4}$ is its cross section area.  These formulas account for the normal axial stress, shear stress, and bending stress, the latter being predominant in this design. 
The factor of safety of the wire ($\mathrm{FoS}_w>$3) can be computed using Eq. (\ref{eqn:wire_FoS}).
	\begin{equation}
	\mathrm{FoS}_w = \frac{\sigma_\mathrm{VM}^w}{\sigma_\mathrm{yield}^w}
	\label{eqn:wire_FoS}
	\end{equation}

\noindent
where $\sigma_\mathrm{yield}$ is the yield strength of the wire, and $\sigma_\mathrm{VM}$ is the equivalent Von Mises stress as defined in Eq. (\ref{eqn:wire_sigma}). 
	\begin{equation}
	\sigma_\mathrm{VM}^w =  \sqrt{\sigma_\mathrm{max}^2+\tau_\mathrm{max}^2}
	\label{eqn:wire_sigma}
	\end{equation}

The wire diameter is small enough to place the violin modes at frequencies much higher than the blade vertical mode ($\sim$763 Hz). The approximate formula for the violin modes (which takes into account some anharmonicity due to the elasticity of the wires) is \cite{violin}:
	\begin{equation}
	\begin{split}
	f_{_{\mathrm{VIOLIN}}} & = \ \sqrt{\frac{T}{\rho_w A_w}}\frac{n \pi}{l_w}  
	\\ &\times\left( 1+\frac{2 z_{_\mathrm{ZMP}}}{l}+ \left(\frac{2 z_{_\mathrm{ZMP}}}{l} \right)^2 + 
	\left(\frac{n \pi z_{_\mathrm{ZMP}}}{l} \right)^2 \right)
	\label{eqn:violin}
	\end{split}
	\end{equation}

\noindent
where $T = mg$ is the tension on the wire, $\rho$ is the density of the wire, and $n$ is the mode number.

\begin{table}[b!]
\caption{\label{tab:Horiz} Flexure parameters}
\centering
\begin{spacing}{1.5}
\begin{tabularx}{.94\columnwidth}{p{55mm}c}
 \hline \hline
 Parameter & Value \\ 
 \hline\hline
 Wire length ($l_w$) & 145.6 $\times 10^{-3} \ \mathrm{m}$  \\
 Wire diameter ($d_w$) & 0.61 $\times 10^{-3} \ \mathrm{m}$ \\  
  \hline
 Horizontal resonance $f_{xx}$ & 1.26 $\mathrm{Hz}$ \\
 Spring constant (single blade) ${k_{xx}}$ & 790.5 $\frac{\mathrm{N}}{\mathrm{m}}$  \\
$z_{_\mathrm{ZMP}}$ & 3.41 $\times 10^{-3} \ \mathrm{m}$  \\
$\mathrm{FoS}_w$ & 3.65 \\
 \hline\hline
\end{tabularx}
\end{spacing}
\end{table}

\subsection{\label{sec:vert_is}Vertical isolation}
The vertical isolation is provided by three triangular stainless steel cantilever blade springs. This shape guarantees a homogeneous stress distribution (assuming infinitesimal strain). The vertical stiffness of the blade ($k_{zz}$) and first resonance of the suspension ($f_{zz}$) can be approximated by Eq. (\ref{eqn:vert_stif}) and Eq. (\ref{eqn:vert_freq}), respectively. 
	\begin{equation}
	k_{zz} = \sqrt{\frac{E_b a_b h_b^3}{6 l_b^3}}
	\label{eqn:vert_stif}
	\end{equation}
	\begin{equation}
	f_{zz} = \frac{1}{2 \pi}\sqrt{\frac{k_{zz}}{m}}
	\label{eqn:vert_freq}
	\end{equation}
	
	\noindent
	where $E_b$ is the blade Young's modulus, $a_b$ is the width of the blade at the base, $h_b$ is the blade thickness, $m$ is the mass supported by each blade, and $l_b$ is the length of the blade. 
	
	Using infinitesimal strain theory, we can estimate the vertical deflection ($\omega_\mathrm{tip}$) and angle ($\omega'_\mathrm{tip}$) at the tip of the blade (Eq. \ref{eqn:vert_tip}), as well as the maximum stress ($\sigma_\mathrm{max}^{b}$) and factor of safety ($\mathrm{FoS}_b$) of the blade (Eq. \ref{eqn:blade_stress}).
	\begin{equation}
	\omega_\mathrm{tip} = \frac{6 m g l_b^3}{E_b a_b h_b^3} \ \qquad    \omega'_\mathrm{tip} = \frac{12 m g l_b^2}{E_b a_b h_b^3}
	\label{eqn:vert_tip}
	\end{equation}
	\begin{equation}
	\sigma_\mathrm{max}^{b} =  \frac{6 m g l_b}{a_b h_b^2}  \ \qquad   \mathrm{FoS}_b = \frac{\sigma_\mathrm{max}^{b}}{\sigma_\mathrm{yield}^{b}}
	\label{eqn:blade_stress}
	\end{equation}

Since there are three design parameters ($a_b$, $l_b$, and $h_b$) and only two constraints ($f_{zz}$ and $\sigma_\mathrm{max}^{b}$ from Eq. (\ref{eqn:vert_freq}) and Eq. (\ref{eqn:blade_stress})), the design has one degree of freedom. We defined the length $l_b$ to be well suited to the overall dimensions of the platform. 

The blades are made from grade 440C stainless steel and are manufactured flat. This reduces the cost and lead-time compared to previous aLIGO blades, made of maraging steel and machined curved \cite{curved_blades}. However, maraging steel is recommended if creep noise or crackle are a concern \cite{blades}. Design values are given in Table \ref{tab:Blades}.

\begin{figure}	[t!]
	\centering
	\includegraphics[width=0.96\columnwidth]{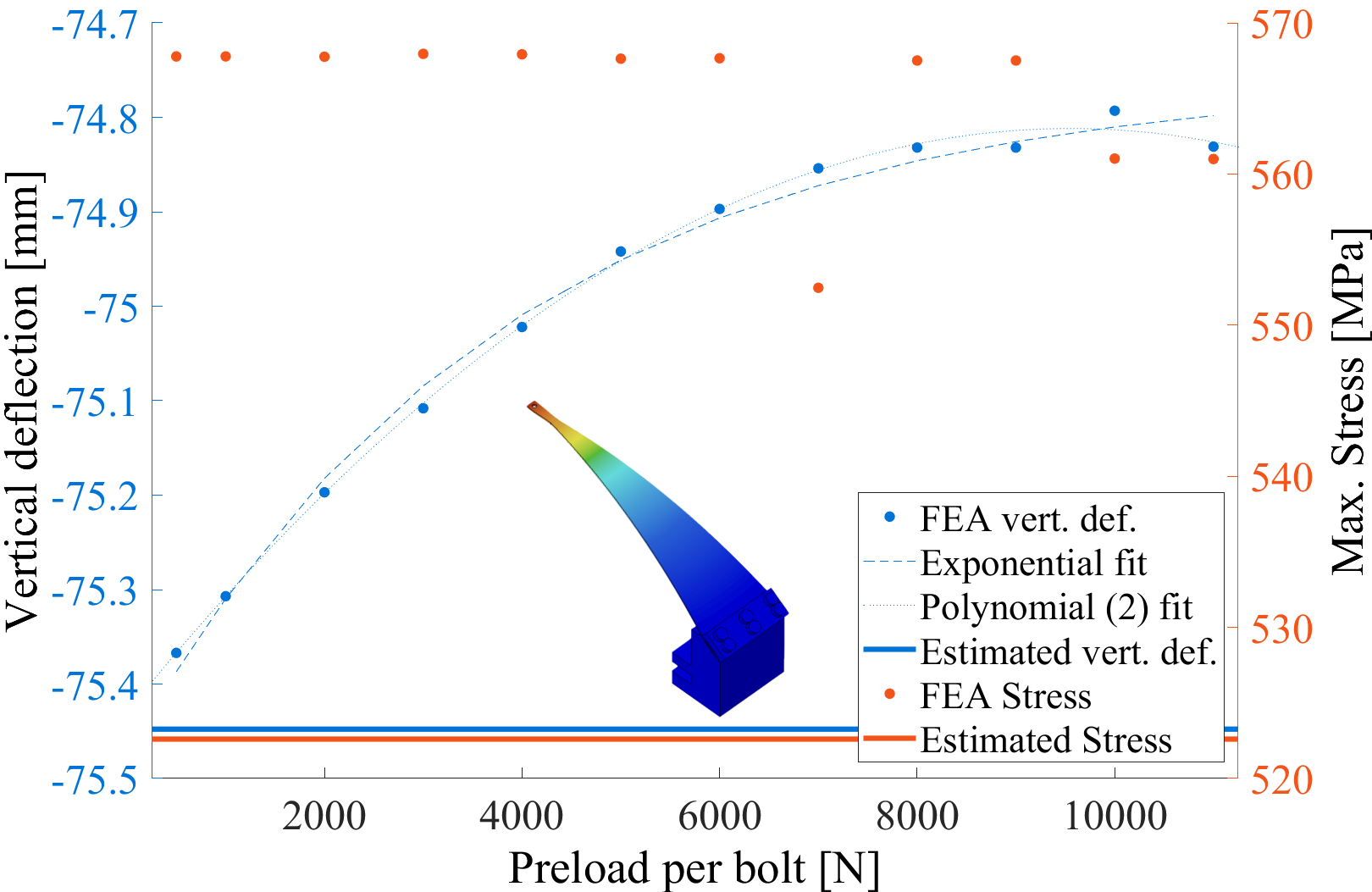}
	\caption{\label{fig:Blade_preload}  \textbf{CLAMP DESIGN.} Results from the FEA for the blade post. The FEA includes the preload in the blade clamp as a parameter and, for each level of preload, the vertical deflection of the blade's tip and the maximum stress in the blade are calculated (points) and compared with their theoretical estimated values (solid lines). This analysis shows the appearance of compliance (more vertical deflection) due to lack of preload. For this specific situation (post pitch angle of 35.32$^\circ$), the minimum amount of preload is $\approx8000$ N. This type of analysis is used to define the right blade post angle and preload at the clamp to ensure that the nominal position of the optical table is reached for the desired mass. }
	\end{figure}

The preliminary design is performed using Eq. (\ref{eqn:vert_freq}-\ref{eqn:blade_stress}), which assumes classical infinitesimal strain beam theory. The blade stiffness and peak stress estimates are verified with finite element analysis.  The relative error between the two calculations is around 1 \% in deflection and 7 \% in stress, as shown in Fig. \ref{fig:Blade_preload}.

\begin{table}[b!]
\caption{\label{tab:Blades} Blade parameters}
\centering
\begin{spacing}{1.5}
\begin{tabularx}{.94\columnwidth}{p{55mm}c}
 \hline \hline
 Parameter & Value \\ 
 \hline\hline
 Design parameters & \\
 Blade base width $a_b$ & 85 $\times 10^{-3} \ \mathrm{m}$ \\ 
 Blade thickness $h_b$ & 2.11 $\times 10^{-3} \ \mathrm{m}$  \\ 
 Blade length $l_b$  & 280 $\times 10^{-3} \ \mathrm{m}$  \\
 Young's modulus $E_b$ & 2.1 $\times 10^{11} \ \mathrm{Pa}$\\
  \hline
 Modeled results &   \\
 Vertical resonance $f_{zz}$ & 1.64 $\mathrm{Hz}$ \\
 Spring constant ${k_{zz}}$ & 1273.1 $\frac{\mathrm{N}}{\mathrm{m}}$  \\
 Tip deflection $\omega_\mathrm{tip}$ & 92.5 $\times 10^{-3} \ \mathrm{m}$  \\
 Tip rotation $\omega'_\mathrm{tip}$ & 0.66 $\mathrm{rad}$ \\ 
$\mathrm{FoS}_b$ & 3.44 \\
 \hline\hline
\end{tabularx}
\end{spacing}
\end{table}

The height and angle of the blade assembly base are determined using FEA to accurately predict the deformation of the blade. Contact analysis was used to define adequate clamping and bolt preload condition (Fig. \ref{fig:Blade_preload}).

Due to its annular shape, stage 0 has low torsional stiffness and relies on being well clamped to the support structure. FEA was used to evaluate the number of clamping elements necessary to establish good contact, and results were verified experimentally.

The rotational stiffness of the isolation system can be estimated using the translation stiffness and the geometrical location of the blade tips as shown in Eq. (\ref{eqn:k_roll}) and Eq. (\ref{eqn:k_yaw}) for an equilateral triangular distribution.

	\begin{equation}
	k_\mathrm{roll} (= k_\mathrm{pitch}) = \frac{3}{2} k_{zz} R^2 + 3 k_{yy} h_{_{_{Z}}}^2 - 3 m g h_{_{_{Z}}}
	\label{eqn:k_roll}
	\end{equation}
	\begin{equation}
	k_\mathrm{yaw} = \frac{3}{2} (k_{xx}+k_{yy}) R^2 + 3  \frac{G_w I_w}{l_w}
	\label{eqn:k_yaw}
	\end{equation}
where $h_{_{_{Z}}} = z_{_\mathrm{CoM}}-z_{_\mathrm{ZMP}}$ is the vertical offset between the center of mass (CoM) and the ZMP, $G_w$ is the shear modulus of the wire, and $R$ is the horizontal distance between the center of gravity and the tip of the blades.

Due to the vertical offset between the CoM and the ZMP ($h_{_{_{Z}}}$), some of the modes are coupled. This leads to off-diagonal terms in the 6x6 stiffness matrix (Eq. \ref{eqn:k_matrix}). 

	\begin{equation}
	\mathbf{K} =  
	\left[
	\begin{matrix}
   	 3 k_{xx} & 0 & \kappa_{xz} & 0 & 3 k_{xx} h_{_{_{Z}}} & 0 \\
   	 0 & 3 k_{yy} & \kappa_{yz} & 3 k_{yy}h_{_{_{Z}}} & 0 & 0  \\
	 \kappa_{xz} & \kappa_{yz} & 3 k_{zz} & 0 & 0 & \kappa_{zrz}  \\
 	 0 & 3 k_{yy} h_{_{_{Z}}} & 0 & 3 k_\mathrm{roll} &  \kappa_{rxry} & 0  \\
 	 3 k_{xx}h_{_{_{Z}}} & 0 & 0 & \kappa_{rxry} & 3 k_\mathrm{pitch} & 0  \\
 	 0 & 0 & \kappa_{zrz} & 0 & 0 & 3 k_\mathrm{yaw}  \\
	\end{matrix}
	\right]
	\label{eqn:k_matrix}
	\end{equation}

Furthermore, the blades are curved when the suspension is loaded (Fig. \ref{fig:Blade_Post}) and their axes do not form an equilateral triangle (Fig. \ref{fig:VOPO_CAD}). These effects also introduce cross-coupling between the modes ($\kappa_{ij}$). Their values can be approximated with an analytic expression, but the derivation is out of the scope of this paper. However, one of the design considerations taken into account is to reduce the vertical offset between the CoM, the ZMP, and the horizontal actuation plane (i.e., the height of the horizontal actuators). Note that the zeros in the stiffness matrix (Eq. \ref{eqn:k_matrix}) represent couplings that are not relevant in our design but that could potentially be relevant for a different blade distribution.

Due to the cross couplings, the resonant frequencies are better approximated solving the eigenvalue problem in Eq. \ref{eqn:eigen}.

\begin{equation}
[\mathbf{M}^{-1}\mathbf{K}] \ \boldsymbol{\phi}_n = -(2\pi f_n)^2 \ \boldsymbol{\phi}_n
\label{eqn:eigen}
\end{equation}
where $\mathbf{M}$ is the mass matrix, $f_n$ is the $n^{th}$ resonant frequency, and $\boldsymbol{\phi}_n$ is the mode shape of that resonance. The results of this calculation are summarized in Table \ref{tab:freq}.

\section{\label{sec:actdamp}Active Damping}
\begin{figure}	[b!]
	\centering
	\includegraphics[width=0.96\columnwidth]{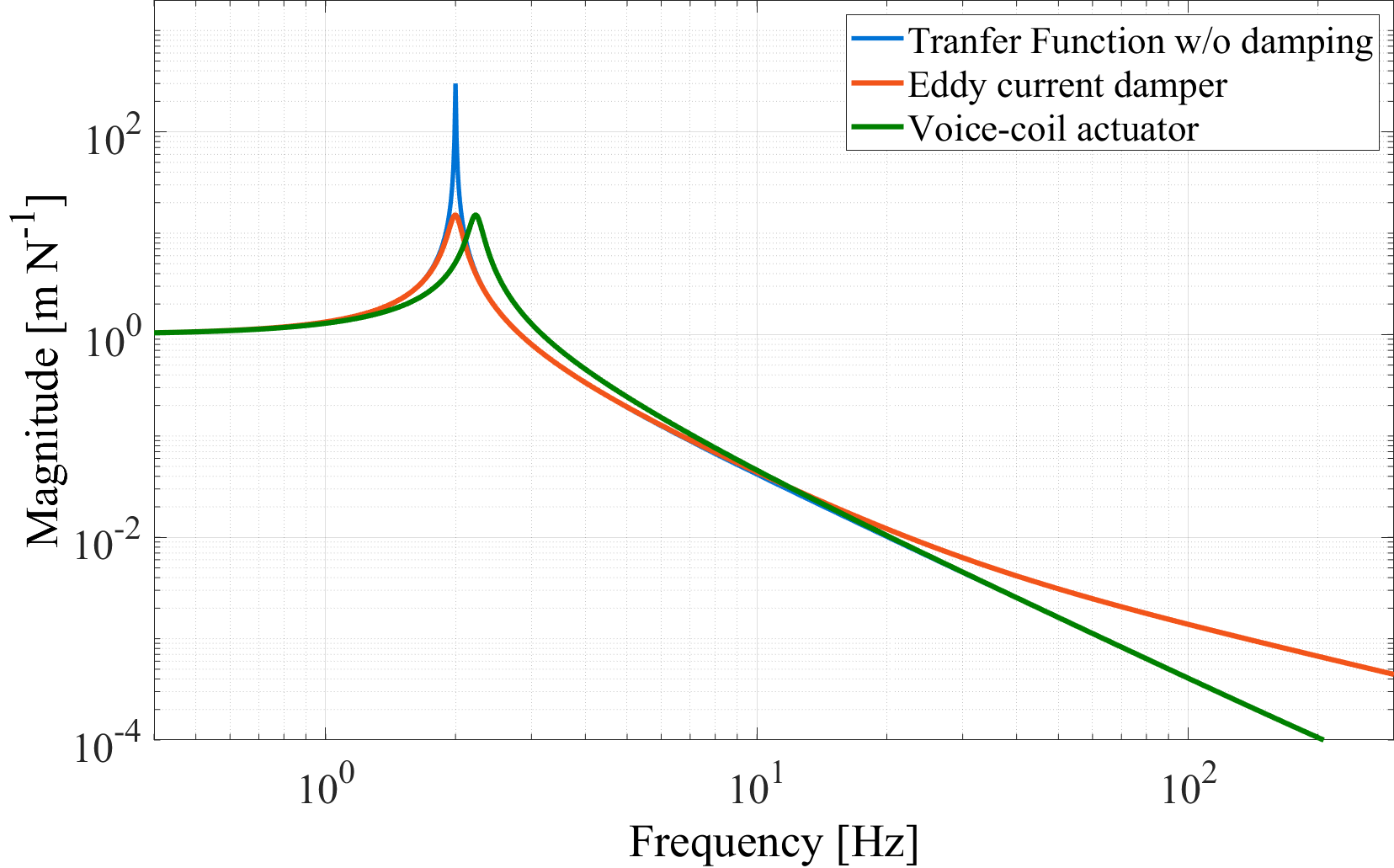}
	\caption{\label{fig:Eddy}  \textbf{DAMPING.} Numerical comparison of the performance of the damping options. The un-damped resonance at 2 Hz has a Q of 300 (typical value for the suspension). Both the passive and active damping reduce the Q to 15. The passive damper impacts the isolation at high frequency ($\approx$ 10 dB at 100 Hz, $\approx$ 20 dB at 300 Hz)`. The active damper uses 2 poles to roll off the viscous coupling at high frequency and preserve the isolation. The figure also shows that the controller might add some stiffness to the system, resulting in a shift in the resonance frequency (see Fig. \ref{fig:Eddy}). Additionally, some gain picking might happen for frequencies above the resonance. This effect is discussed in Sec. \ref{subsec:active}}
	\end{figure}

The rigid body modes of the suspension must be damped to avoid excess motion at the resonance frequencies. This can be done effectively using passive actuators such as eddy current dampers, at the cost of compromising the passive isolation at high frequency through viscous coupling. This is illustrated by the transfer functions in Fig. \ref{fig:Eddy}, where the passive damping reduces the resonance's quality factor Q to value of 15, while increasing the motion by 10 dB at 100 Hz.
To alleviate this loss of isolation, active voice-coil damping has been chosen for this system (Fig. \ref{fig:Eddy}). Note that, in this plot, the effect of sensor and actuator noise was not taken into account. Their effect is discussed in Sec. \ref{subsec:scheme}. The presented example does show the effects that adding roll-off in the controller might have. If the roll-off begins close to the resonance, the control loop will add some stiffness to the system, resulting in a shift in the resonance frequency (see Fig. \ref{fig:Eddy}), which results in gain-picking above the resonance. Nevertheless, these effects are specific to the type of controller implemented in the system. Therefore a compromise has to be found between stiffening the system, gain picking and high frequency noise injection. This topic is further discussed in  Sec. \ref{subsec:active}.

The LIGO Scientific Collaboration (LSC) has produced two types of low-noise ultra-high vacuum compatible collocated actuator-sensor pairs, named AOSEM and BOSEM \cite{aston2011optical,BOSEM}. Their actuation strengths (for a 2x6 mm SmCo magnet) are 0.0309 N/A and 1.694 N/A, respectively. For the test results presented in the following sections, the platforms were equipped with AOSEMs. In future applications, they can be equipped with BOSEMs to provide additional steering range.

\subsection{Sensor-actuator pairs}
  
\begin{figure}	[t!]
	\centering
	\includegraphics[width=0.96\columnwidth]{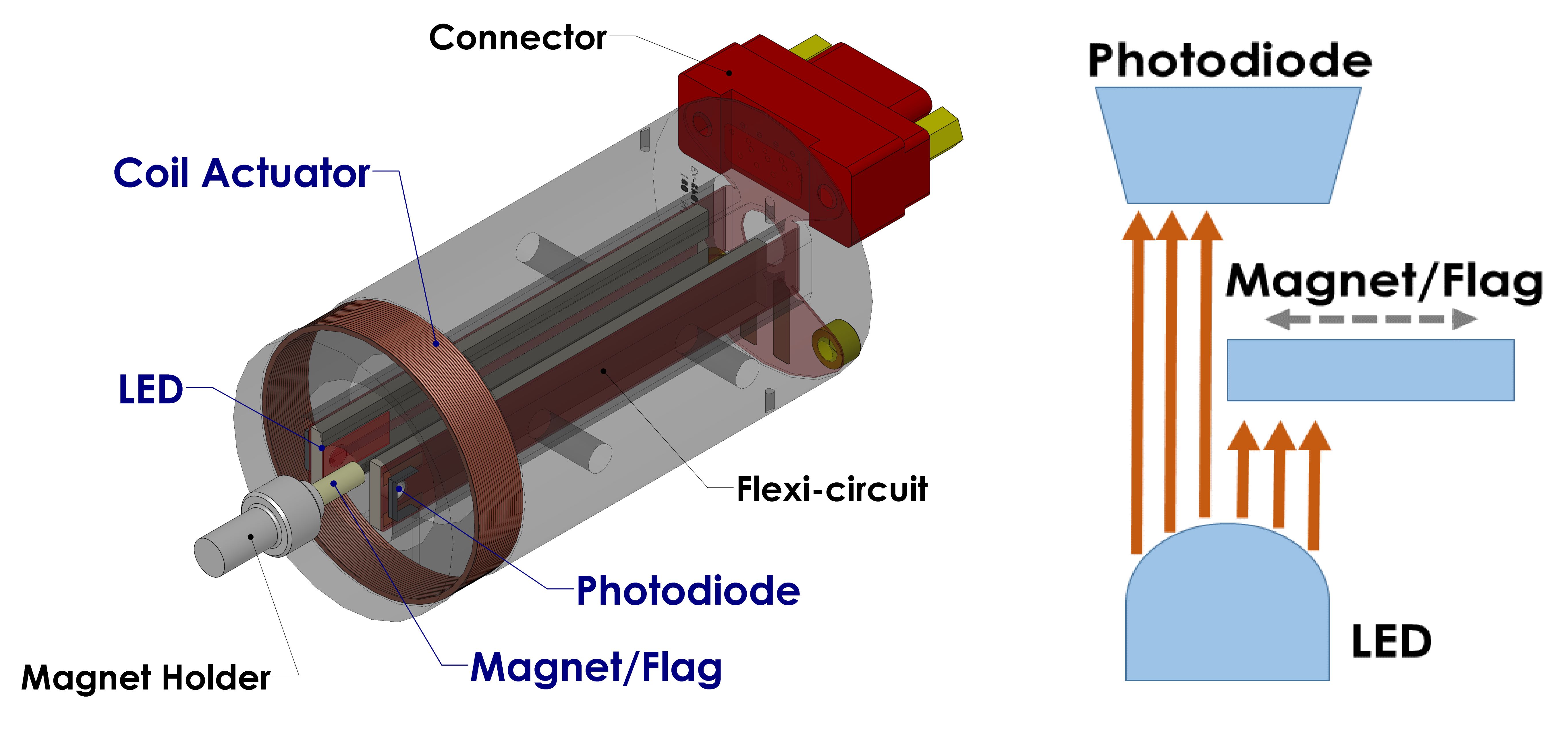}
	\caption{\label{fig:AOSEM} \textbf{LOW NOISE, UHV COMPATIBLE SENSOR-ACTUATOR.} AOSEM schematics. Motion is sensed with a LED-PhotoDiode shadow sensor. At the nominal position, the flag (magnet) covers the shadow sensor at 50\% of its full range. The actuation is provided by the coil-magnet pair. An electrical current through the AOSEM's coil generates a magnetic field, which acts on the magnet.}
	\end{figure}

Fig. \ref{fig:AOSEM} shows the sensor-actuator's basic principle, where a nickel plated SmCo magnet acts both as a flag for the shadow sensor and as the actuation element within the coil. An electrical current in each voice-coil generates a magnetic field, which acts on the magnets attached to stage 1. The isolation platform is equipped with six sensor-actuators, grouped in horizontal-vertical pairs as shown in Fig. \ref{fig:VOPO_CAD}, thus damping all six rigid body modes. 

The operation point is such that the flag covers 50\% of the range of the sensor when stage 1 is floating at its equilibrium position. In order to prevent sensor damage, the suspension is equipped with hard stops, limiting the translation to less than 1mm (see Fig. \ref{fig:Blade_Post}).

\subsection{Control scheme and noise budget} \label{subsec:scheme}
\begin{figure}	[b!]
	\centering
	\includegraphics[width=0.96\columnwidth]{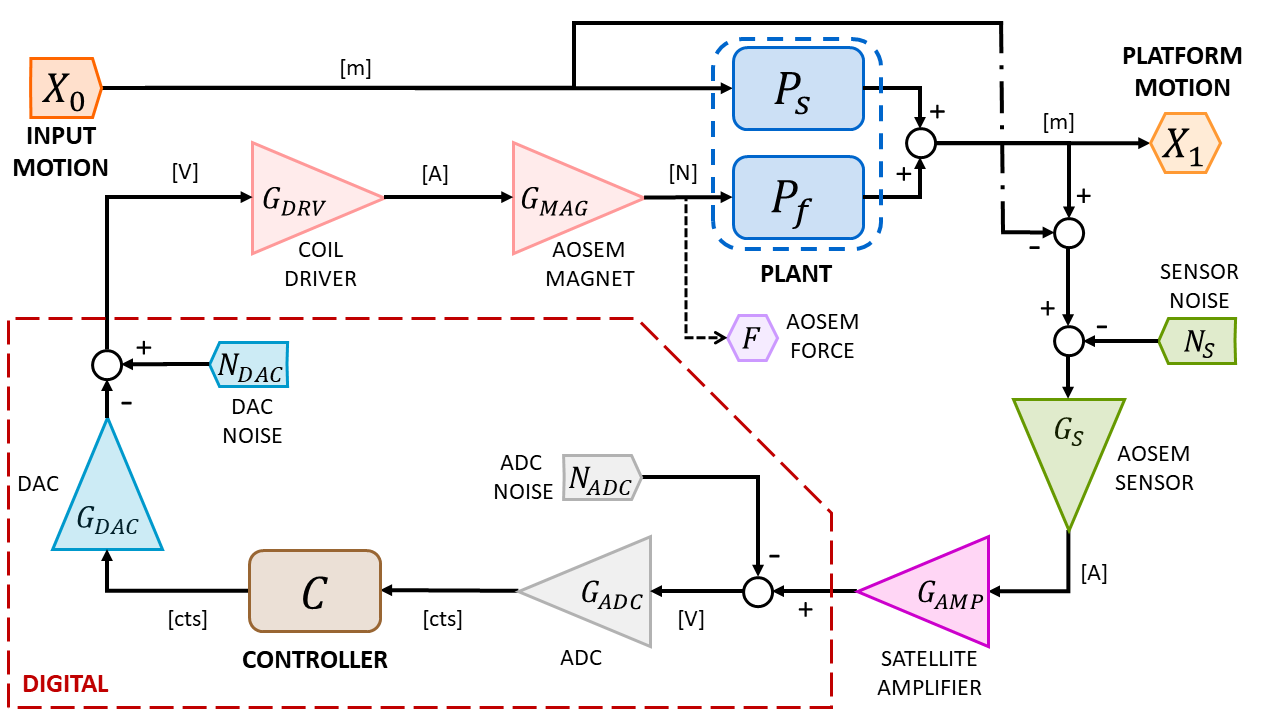}
	\caption{\label{fig:Cont_schme}\textbf{CONTROL DIAGRAM.}  Control diagram of the isolation platform. The analog signal from the AOSEMs is digitized before to be sent to the controller, which generates a signal that is converted and sent to the AOSEMs' coil driver. Besides seismic noise, this diagram introduces control noise due to the sensor, the analog-to-digital (ADC), and digital-to-analog (DAC) converters.}
	\end{figure}

Fig. \ref{fig:Cont_schme} shows the control diagram and the noise sources, namely the input motion and the control noise. The latter includes sensor, analog-to-digital (ADC), and digital-to-analog (DAC) converter noises. A simplified version of the seismic path ($P_s$) and the force path  ($P_f$) in the $s$-domain are presented in Eq. (\ref{eqn:Ps}) and Eq. (\ref{eqn:Pf}). It does not account for possible cross-couplings between the degrees of freedom, which are negligible for the purpose of this discussion.
	\begin{equation}
	P_{s} = \frac{c s + k}{m s^2 + c s + k}
	\label{eqn:Ps}
	\end{equation}
	\begin{equation}
	P_{f} = \frac{1}{m s^2 + c s + k}
	\label{eqn:Pf}
	\end{equation}
	
\noindent
where $s$ is the Laplace variable. These equations are analogous to a single damped spring-mass system of mass $m$, stiffness $k$, and Q-factor $Q = \frac{1}{2\zeta} = \frac{\sqrt{km}}{c}$.
	
Given this control diagram, the input motion ($X_0$), the sensor noise ($N_{s}$) \cite{sensor}, and ADC/DAC noises ($N_\mathrm{DAC}$, $N_\mathrm{ADC}$) \cite{adc,dac}, we can estimate the platform motion ($X_1$) as it follows:
	\begin{equation}
	X_1 = P_f F + P_s X_0
	\label{eqn:X1}
	\end{equation}

\noindent
where $F$ is the force applied by the actuators, which can be calculated using the gains defined in Fig. \ref{fig:Cont_schme} as:
	\begin{equation}
	\begin{split}
	F & = \big[G_{_\mathrm{MAG}} G_{_\mathrm{DRV}} \big] N_{_\mathrm{DAC}} \\ &+ \big[G_{_\mathrm{MAG}} G_{_\mathrm{DRV}}G_{_\mathrm{DAC}} C G_{_\mathrm{ADC}} \big] N_{_\mathrm{ADC}} \\ &+ \big[G_{_\mathrm{MAG}} G_{_\mathrm{DRV}}G_{_\mathrm{DAC}} C G_{_\mathrm{ADC}}G_{_\mathrm{AMP}} G_{_{S}} \big] \big(N_{_{S}} + X_0 - X_1 \big) 
	\label{eqn:F}
	\end{split}
	\end{equation}

\begin{figure}

      \subfloat[\label{fig:ground_vs_HAM}]{\includegraphics[width=0.96\linewidth]{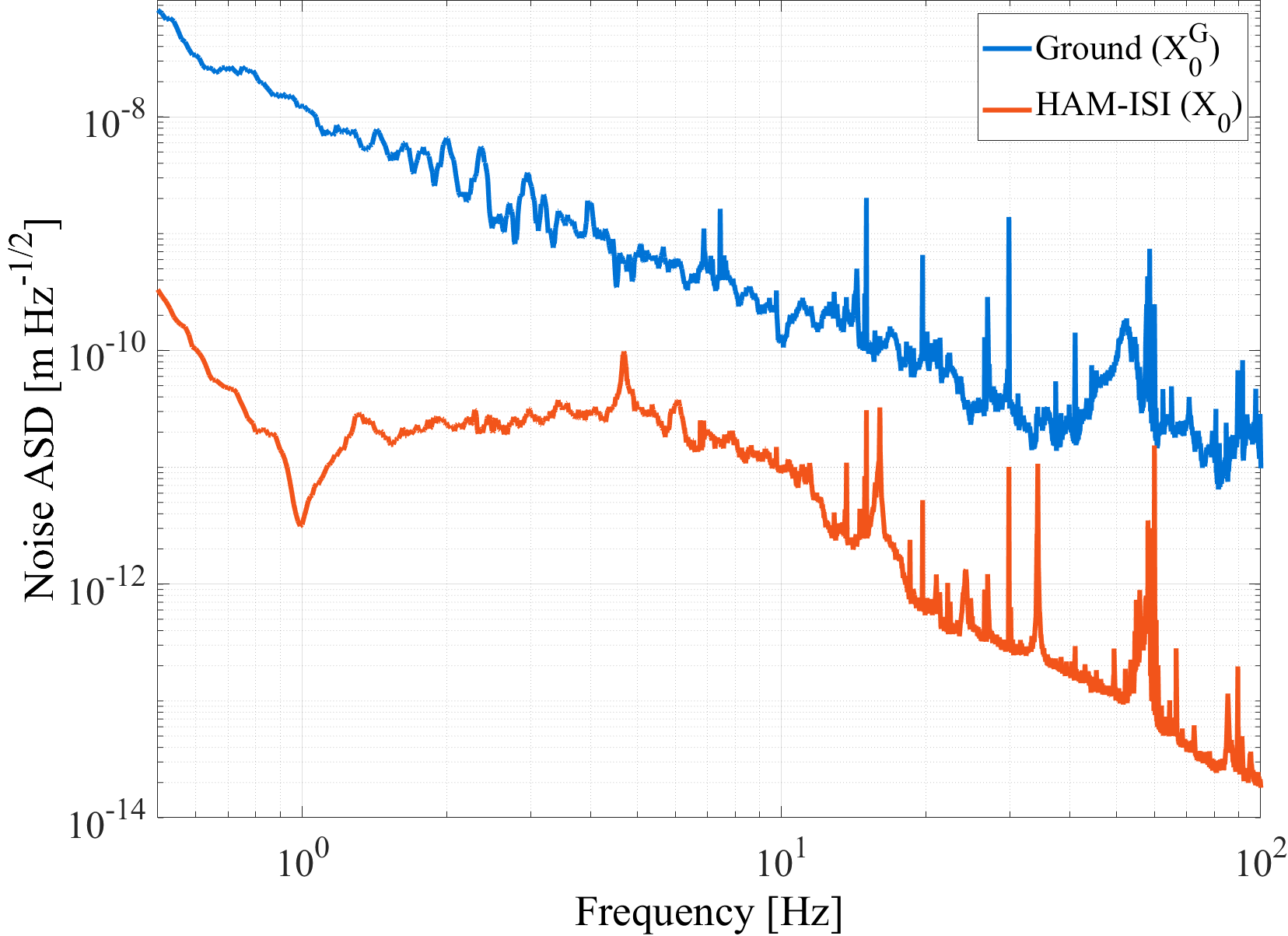}}

      \subfloat[\label{fig:noise_electric}]{\includegraphics[width=0.96\linewidth]{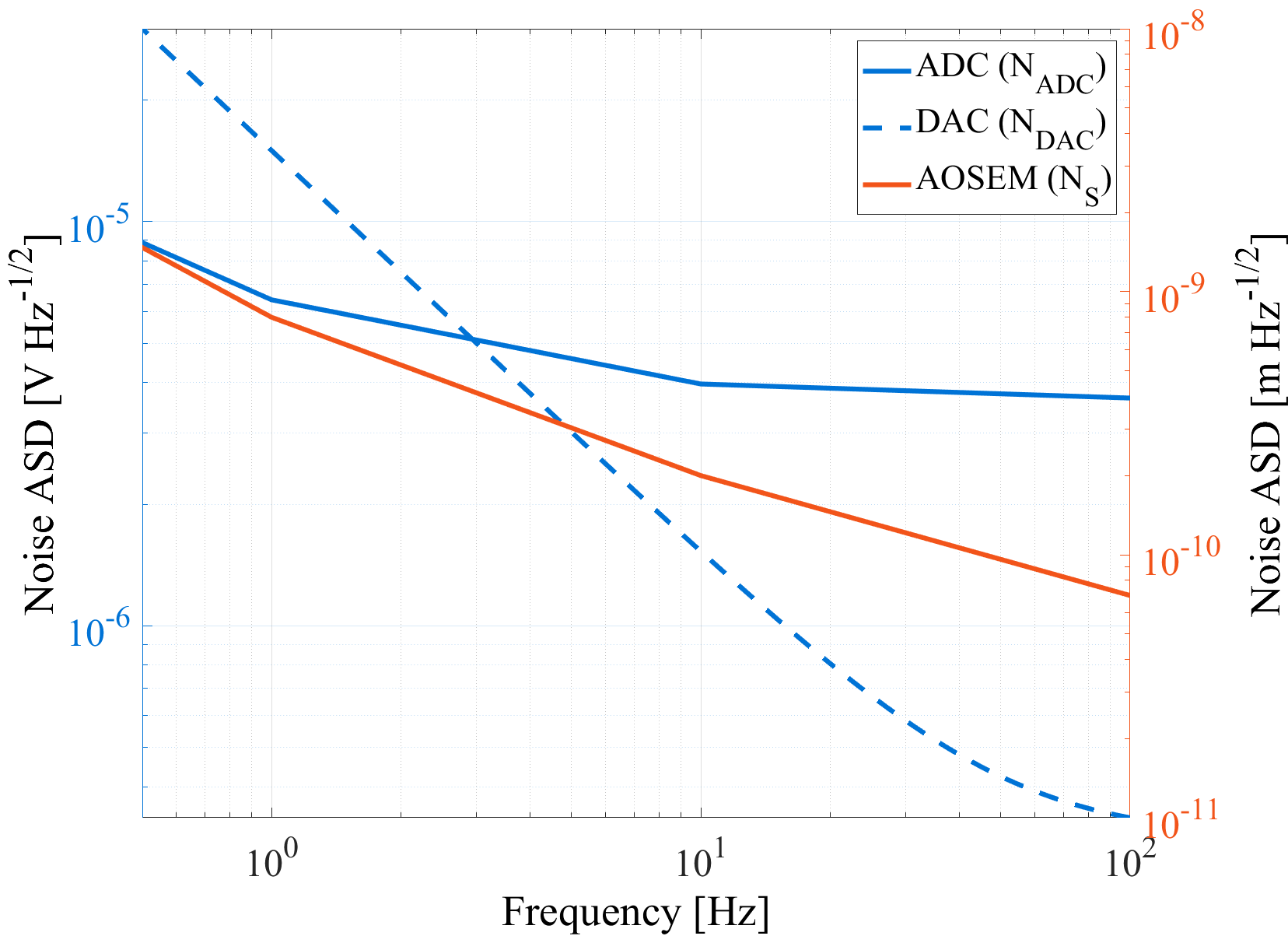}}

   \caption{\label{fig:noise_base}\textbf{INPUT NOISE.} The platform is to be mounted on an isolated HAM-ISI table \cite{HAMISI1}. Thus input motion of our system ($X_0$) is the motion of the HAM-ISI. (a) Compares the Vertical (Z) motion of the HAM-ISI with the typical ground motion ($X_0^G$). The HAM-ISI provides a very low input motion. (b) Amplitude spectral density of sensor, ADC and DAC noise.}	
\end{figure}

	Given that the noise sources are incoherent, we can estimate the total noise using their root mean square (RMS):
	\begin{eqnarray}
	|X_1|^2 \sim |X_{X_0}|^2  + |X_{N_\mathrm{CONT}}|^2 
	\label{eqn:X_1}
	\end{eqnarray}
	where the platform's motion due to ground seismic motion can be estimated as:
		\begin{equation}
	X_{{X_0}} = \frac{\mathrm{RR} + P_s}{1+\mathrm{RR}} \ X_{0}
	\label{eqn:X0}
	\end{equation}
	where $\mathrm{RR}$ is the return ratio (i.e., the open loop gain):
	\begin{equation}
	\mathrm{RR} = P_f G_c 
	\end{equation}
	
	Defining $G_F = G_{_\mathrm{MAG}} G_{_\mathrm{DRV}}$, $G_{\mathrm{AS}} = G_{_\mathrm{AMP}} G_{_{S}}$, and $G_{\mathrm{AD}} = G_{_\mathrm{DAC}} G_{_\mathrm{ADC}}$, we have that:
	\begin{equation}
	G_c = G_F G_{\mathrm{AS}} G_{\mathrm{AD}} C 
	\end{equation}
	
	For a given controller, the total control noise corresponds to:
	\begin{eqnarray}
	|X_{N_\mathrm{CONT}}|^2 \sim |X_{N_\mathrm{DAC}}|^2 + |X_{N_\mathrm{ADC}}|^2 +|X_{N_{S}}|^2 
	\label{eqn:X_elec}
	\end{eqnarray}
where
	\begin{equation}
	X_{N_\mathrm{DAC}} = \frac{P_f G_F}{1+\mathrm{RR}} \ N_{_\mathrm{DAC}}
	\label{eqn:Ndac}
	\end{equation}
	\begin{equation}
	X_{N_\mathrm{ADC}} = \frac{1}{G_{\mathrm{AS}}} \frac{P_f G_C}{1+\mathrm{RR}} \ N_{_{ADC}}
	\label{eqn:Nadc}
	\end{equation}
	\begin{equation}
	X_{N_{S}} = \frac{P_f G_C}{1+\mathrm{RR}} \ N_{_{S}}
	\label{eqn:Ns}
	\end{equation}

	The contribution of these terms to the closed loop motion is quantified and discussed in Sec. \ref{subsec:active}.
\section{\label{sec:perf} Experimental Performance}
A prototype of the system was tested at the MIT-LIGO facilities and two units have been installed at the LIGO observatories in Livingston, LA (LLO) and Hanford, WA (LHO). The results presented in this section are from testing these three units.

\begin{table}[b!]
\caption{\label{tab:freq} Rigid body modes}
\centering
\begin{spacing}{1.5}
\begin{tabularx}{0.94 \columnwidth}{p{33mm}c  c}
 \hline\hline
 Degree-of-freedom &  Measured (Model) & Q-factor \\ 
 \hline
 Longitudinal-$X$ & 1.27 (1.26) $\mathrm{Hz}$  & 260\\ 
 Transverse-$Y$  & 1.27 (1.28) $\mathrm{Hz}$ & 310  \\
 Vertical-$Z$ & 1.74 (1.64) $\mathrm{Hz}$ & 250  \\ 
 Roll-$RX$  & 2.39 (2.23) $\mathrm{Hz}$  & 170 \\
 Pitch-$RY$ & 1.63 (1.56) $\mathrm{Hz}$ & 200 \\
 Yaw-$RZ$ & 1.44 (1.44) $\mathrm{Hz}$  & 250 \\
 \hline\hline
\end{tabularx}
\end{spacing}
\end{table}

\subsection{Transfer functions}
\begin{figure}[t!]
	\centering
	\includegraphics[width=0.96\columnwidth]{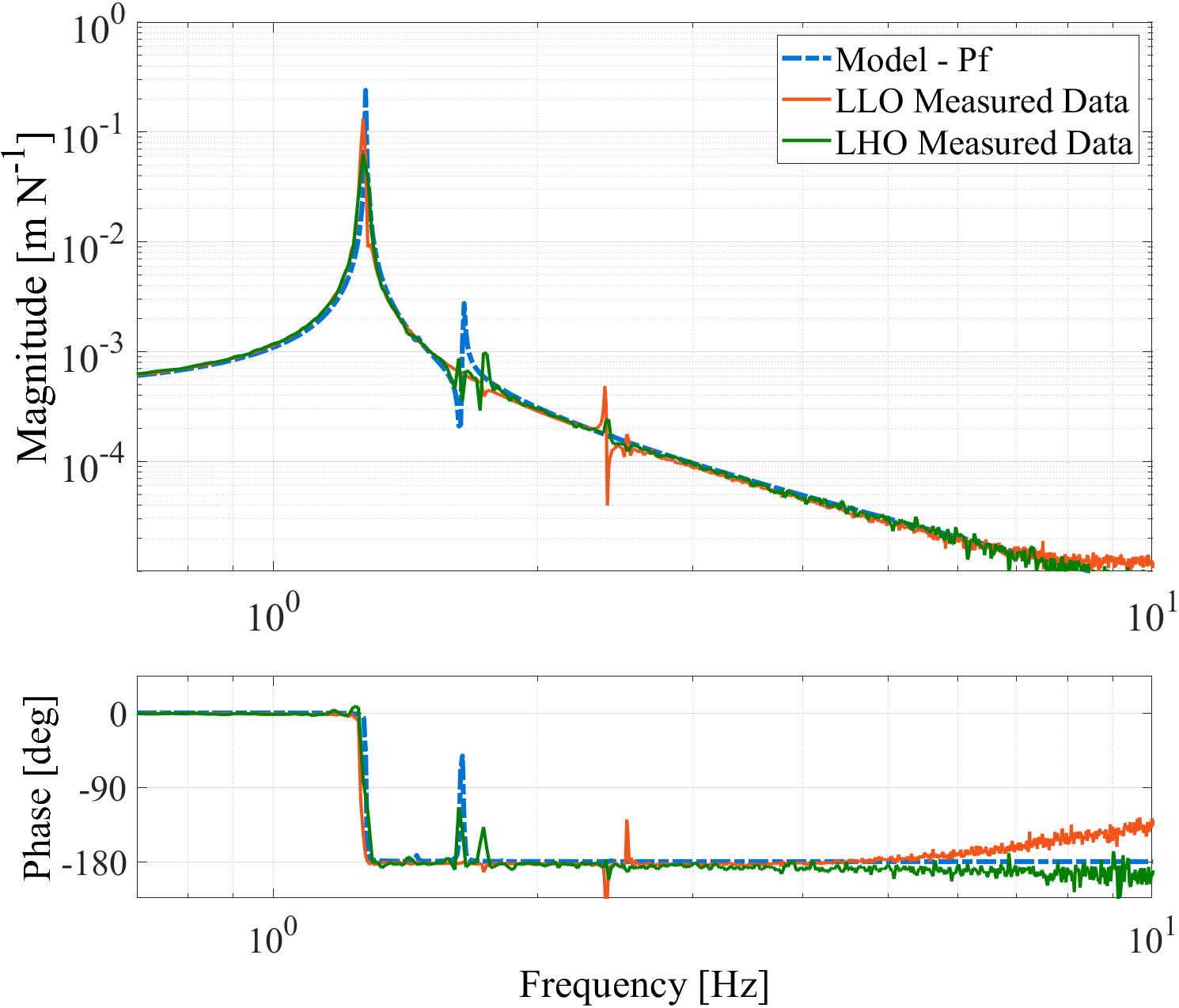}
	\caption{\label{fig:TF} \textbf{TRANSFER FUNCTIONS} Longitudinal (X) transfer function measured at LLO and LHO, DC calibrated to match the dynamical model. The data shows $f^2$ isolation above the resonance. }
	\end{figure}

The transfer functions were obtained by driving the coils and reading out the shadow sensors of the AOSEMs. An example of the suspension's transfer function is presented in Fig. \ref{fig:TF}, which shows good agreement between the model from Eq. (\ref{eqn:k_matrix}) and the measured data. The effect of the cross-coupling terms mentioned in Sec. \ref{sec:vert_is} is visible in this figure. Rigid body mode frequencies and Q-factors for all the degrees of freedom are summarized in Table \ref{tab:freq}.

\subsection{Structural modes}

	\begin{figure}[b!]
	\centering
	\includegraphics[width=0.96\columnwidth]{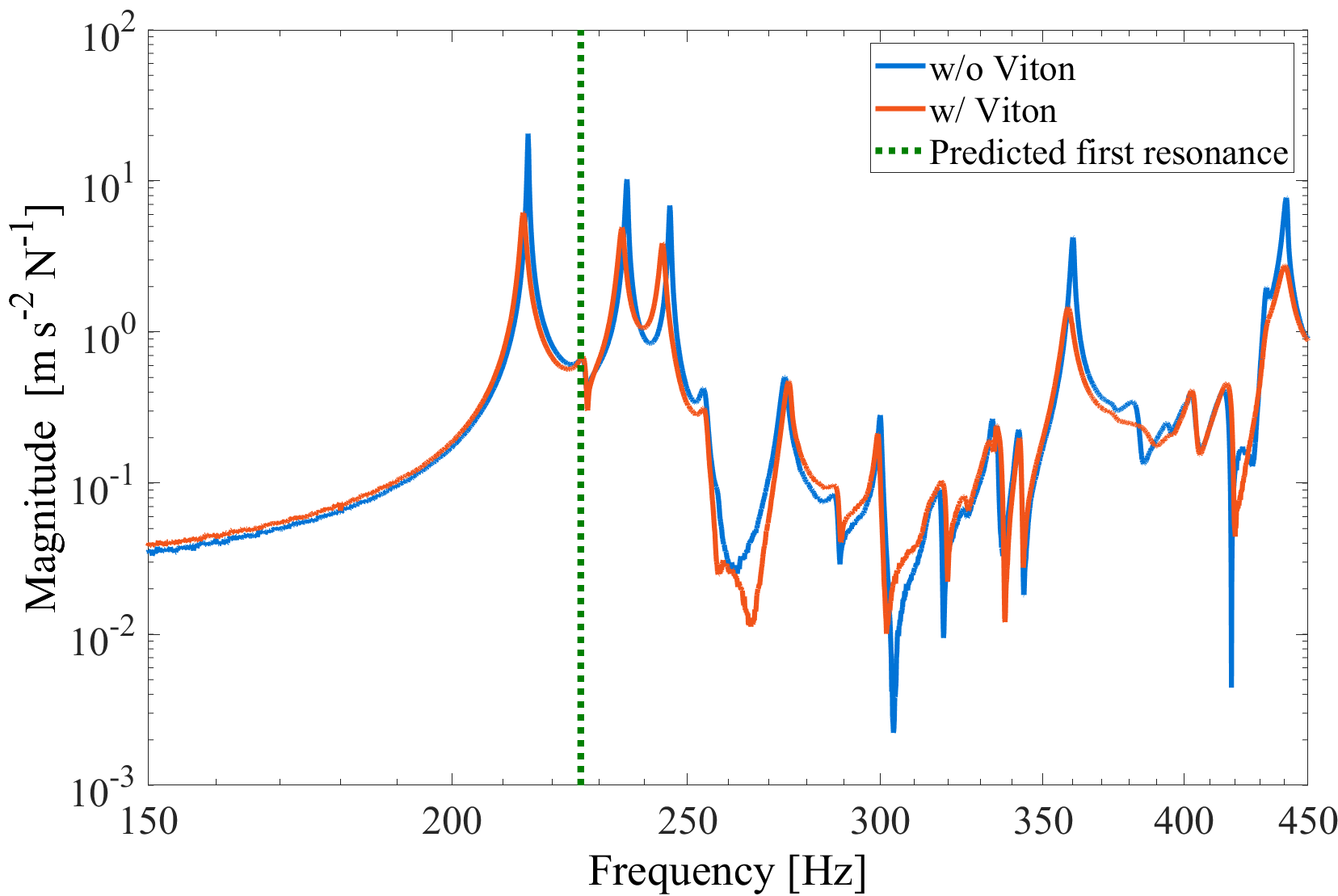}
	\caption{\label{fig:Modal}\textbf{STRUCTURAL MODES.}  Modal analysis results for the vertical (Z) degree of freedom. The Viton pads reduce the mode Qs by about a factor of 2-3. The first resonance is slightly lower than predicted (because the final optics configuration was different than the model) but it is still higher than the required 200 Hz.}
	\end{figure}

The control loops and performance of the HAM-ISI system on which the suspension platform is mounted can be deteriorated by low frequency structural modes. The light and low-profile structure of the suspension minimizes the dynamical impact on the supporting table. Experimental modal analysis of the optical table is consistent with the FEA predictions, as shown in Fig. \ref{fig:Modal}. This response was obtained using a B\&K 8206 impact hammer and a miniature tri-axial accelerometer type 4506 B on the fully loaded suspension. The measured resonances are consistent with predictions for the structural resonances of the elements of the suspension. In order to damp the main structural modes, constrained layer dampers (i.e., Viton pads) were introduced between the balancing masses and the optical table. Fig.  \ref{fig:Modal} also shows the effect of those dampers, reducing the Q factors by about a factor of 2 to 3.

\subsection{\label{subsec:active}Active Damping}
The damping is performed in the Cartesian basis. For that, a change of basis is performed in real time to combine the six sensors and actuators using matrix transformations \cite{matichard2015advanced2}. 

\begin{figure}

      \subfloat[\label{fig:Damp_TF}]{\includegraphics[width=0.96\linewidth]{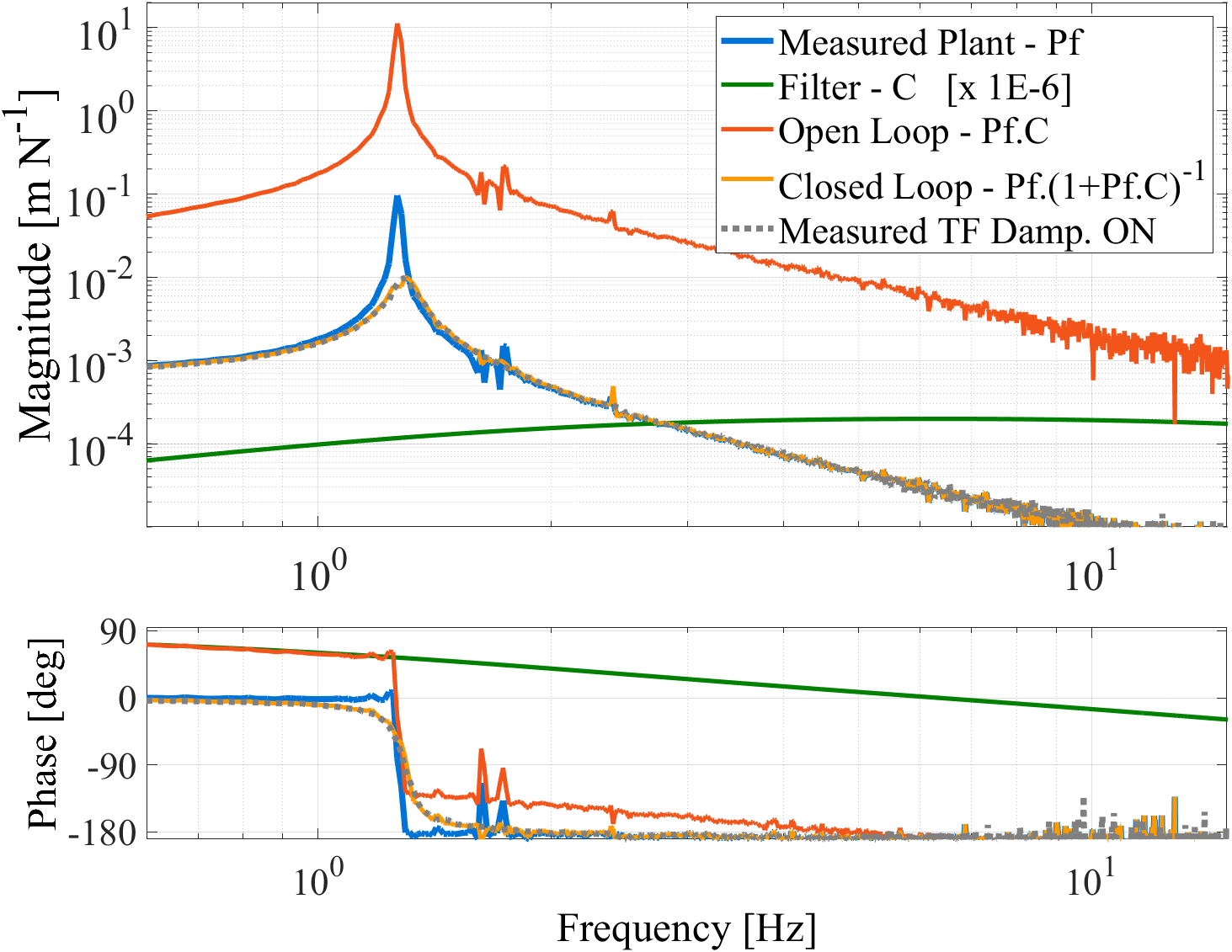}}
      
      \subfloat[\label{fig:PS_long}]{\includegraphics[width=0.96\linewidth]{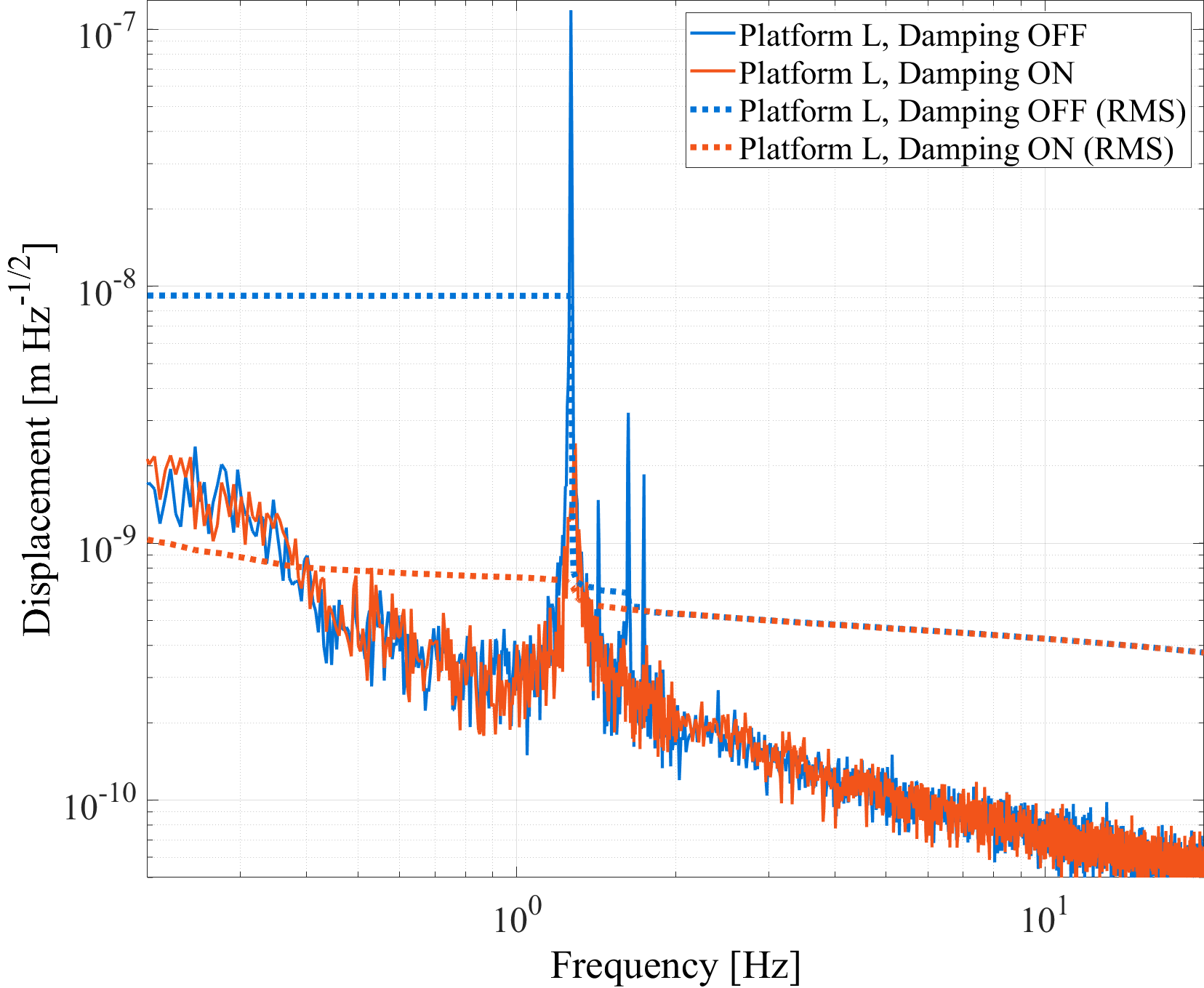}}
      
  \caption{\label{fig:controls}\textbf{ACTIVE DAMPING.} (a) Damping control for the longitudinal (X) degree of freedom. Using a band-pass filter (green), the Q is reduced from $\approx$260 (blue) to $\approx$15 (orange). The phase margin of the controller is $\approx$53$^\circ$ (red). The damped transfer function was measured  (grey) and matches the predicted closed loop. (b) Amplitude spectrale density of the relative displacement ($X_1-X_0$) when the damping loops are ON (red) and OFF (blue) for the longitudinal (X) degree of freedom. The damping provides one order of magnitude of attenuation in RMS relative motion.} 
\end{figure}

The controller (C) is a band-pass filter of the form: 

	\begin{equation}
	C(s) = K \frac{s}{\left( \frac{s}{2\pi f_1} + 1 \right) \left( \frac{s}{2\pi f_2} + 1 \right)} 
	\end{equation}
with one zero at DC and two poles at frequencies $f_1$ and $f_2$. It provides velocity damping at the resonance and limits the injection of control noise off resonance. The controller parameters are tuned to bring the Q values down to about 20, as shown in Fig. \ref{fig:Damp_TF}. Due the filter poles, the controller does not act as a perfect velocity damper at the resonance (controller phase at platform's resonance $\approx 60^{\circ}$), which explains the slight closed loop resonance shift to higher frequency. The corresponding motion amplification at the unity gain frequency (gain-picking) is negligible as will be shown in the next sections.

Fig. \ref{fig:controls} shows the motion of the platform with and without active damping. The control loops reduce the relative sensor RMS motion by an order of magnitude. The motion amplification off resonance due to electronic and sensor noise is quantified by noise budgeting.

\subsubsection{\label{subsec:noise_bdgt}Noise budget}
\begin{figure} 
 \subfloat[\label{fig:noise_elec}]{\includegraphics[width=0.96\linewidth]{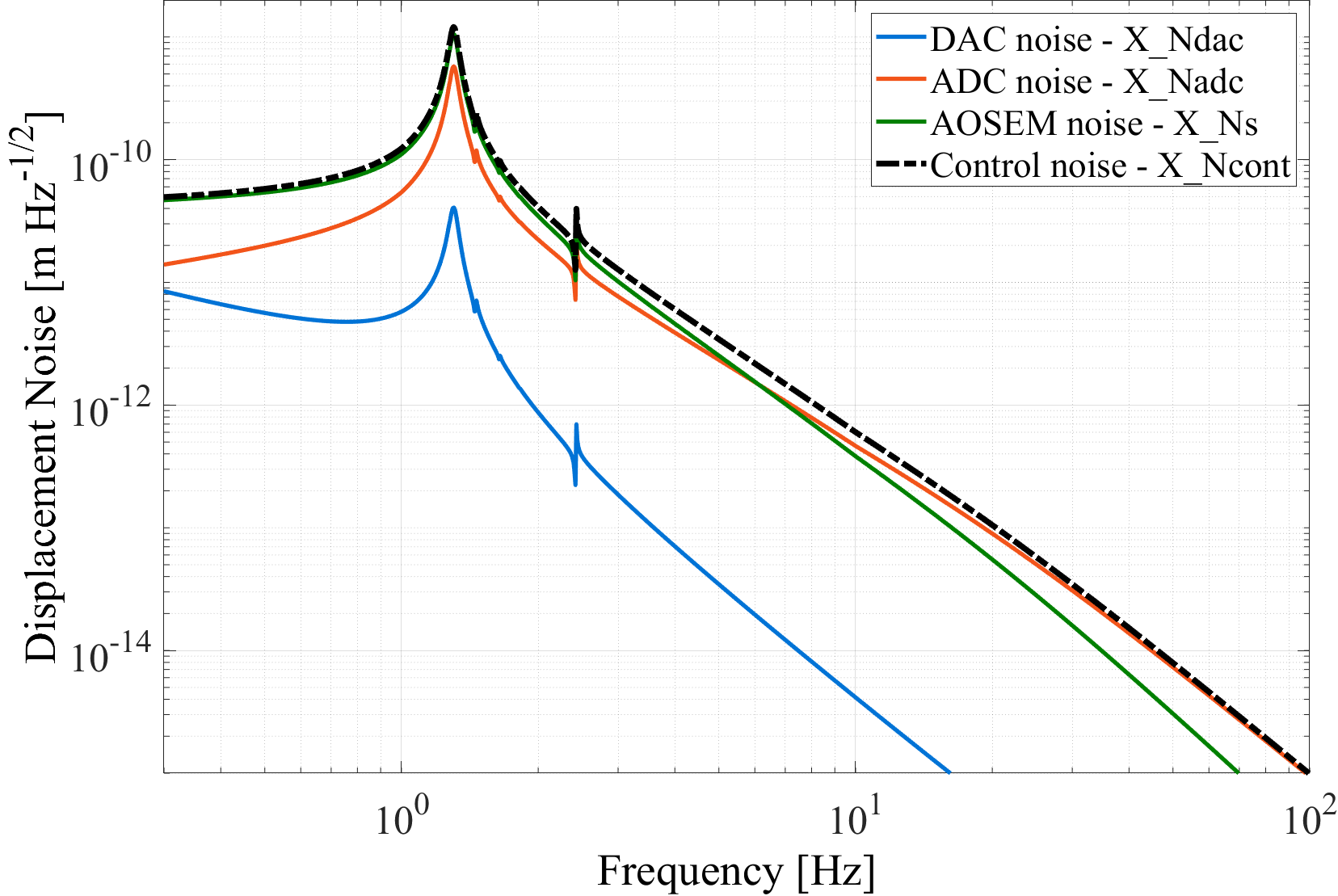}}

\subfloat[\label{fig:noise_total}]{\includegraphics[width=0.96\linewidth]{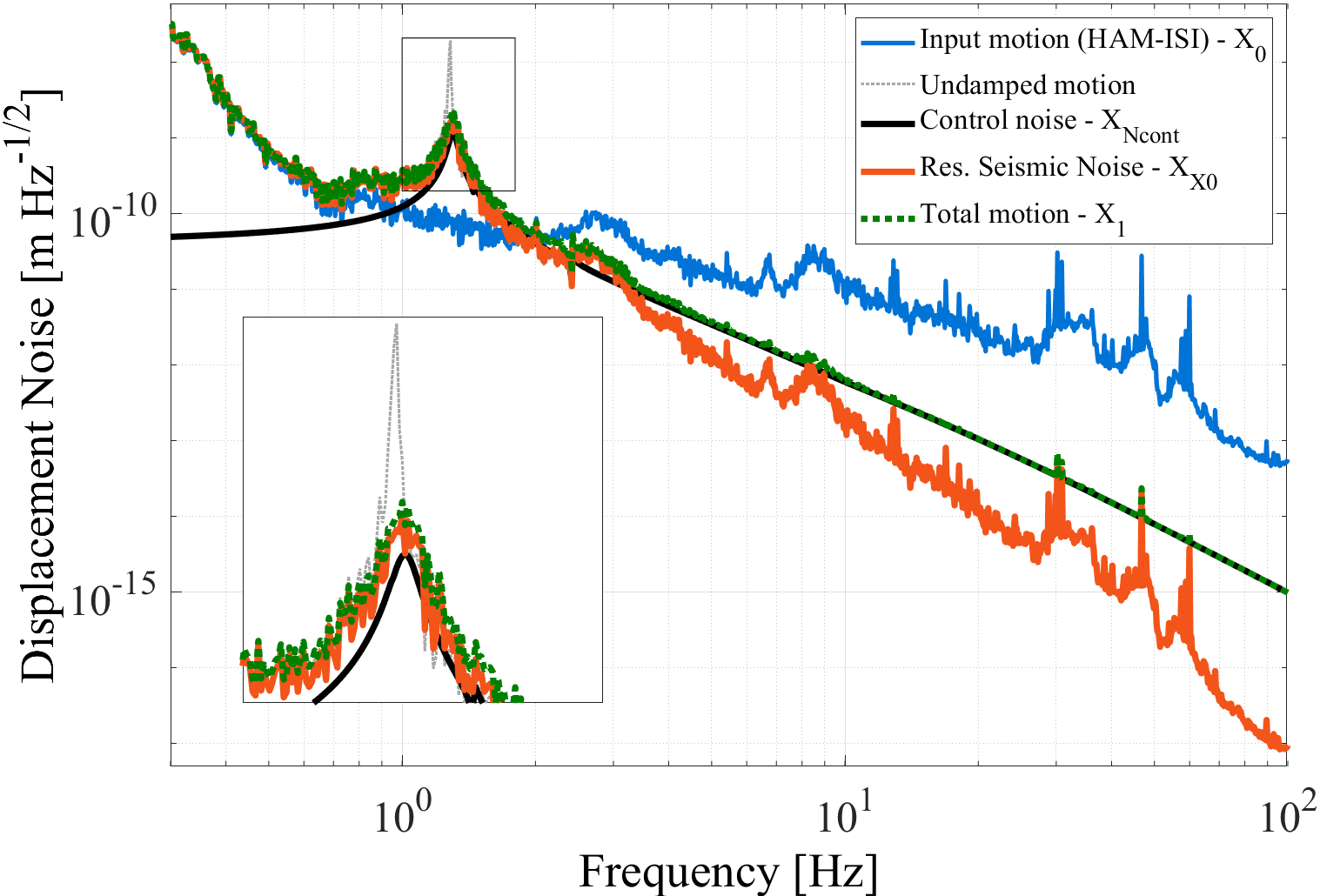}}

  \caption{\label{fig:noise} \textbf{NOISE BUDGET.} Noise budget for the transverse (Y) degree of freedom. (a) Estimate of the platform motion due to each noise source, using the platform's modelled transfer function. At low frequencies, the sensor noise predominates, whereas high frequencies are dominated by ADC noise. (b) The motion produced by the control noise (dashed black) is compared to the estimated motion produced by the input seismic motion (dashed orange). In this example, the controls are the main source of platform motion above 3 Hz. This figure also shows a comparison between the HAM-ISI motion ($X_0$) and the total motion of the suspended stage ($X_1$). }
\end{figure}

The goal of the noise budget is to estimate the control noise injected into the platform and the corresponding motion induced, as shown in Fig. \ref{fig:noise}. Firstly, Fig. \ref{fig:noise_elec} shows that the platform motion due to the control electronics noise is due predominantly to sensor noise at low frequencies and ADC noise at high frequencies. Fig. \ref{fig:noise_total} also compares the motion of the platform ($X_1$, dashed green) with the input motion at its base ($X_0$, blue), showing explicitly the effect of the presented isolator, amplifying the motion around the frequency and reducing it at higher frequencies. 
However, Fig. \ref{fig:noise_total} also compares the estimated platform motion due to the controls noise ($X_{N_\mathrm{CONT}}$, black) and the one due to the input motion ($X_{X_0}$, orange). Due to the low input motion from the HAM-ISI (see Fig. \ref{fig:ground_vs_HAM}), and despite the fact of using low-noise sensors and electronics (see Fig. \ref{fig:noise_electric}), the control noise preponderates above 3 Hz. This results in a deterioration of the isolation at high frequencies. Indeed, at 100 Hz, the displacement noise caused by the controls is $\approx$40 dB higher than the one caused by the input motion, resulting in only $\approx$25 dB of isolation at that frequency. 

\subsubsection{\label{subsec:HF_contr}High-frequency controller tuning}
 The baseline band-pass controller can be tuned to reduce the injection of control noise off resonance. It can be done by either adding extra roll-off at high frequency (poles locations) or by reducing the overall gain of the control loop (or both). Rolling off the controller affects the phase margin, while reducing the overall gain decreases the damping, resulting in larger motion at the resonance. 

\begin{figure}[t!]

\subfloat[\label{fig:controller_comparison}]{%
  \includegraphics[clip,width=0.96\columnwidth]{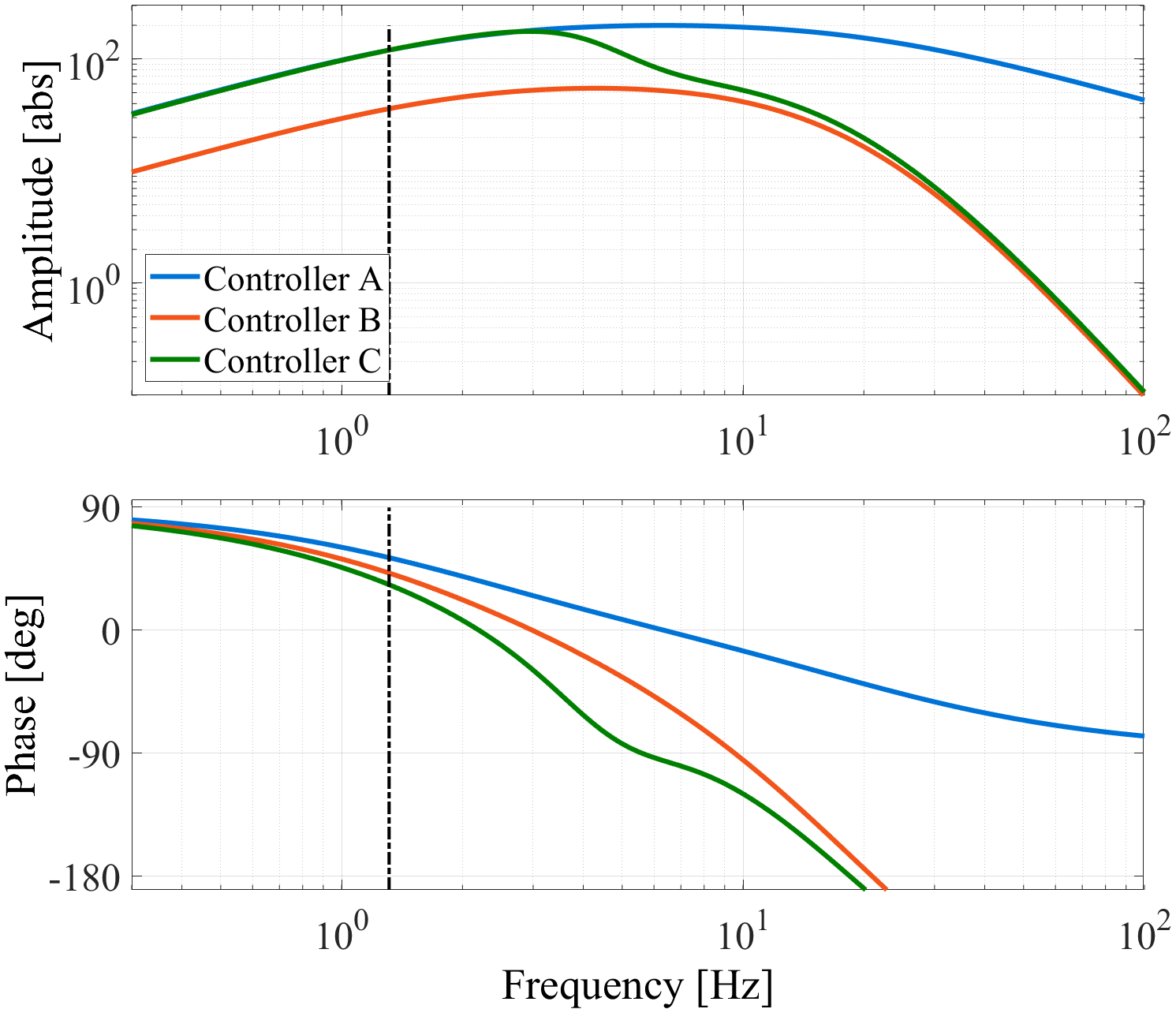}
}

\subfloat[\label{fig:controller_comparison_noise}]{%
  \includegraphics[clip,width=0.96\columnwidth]{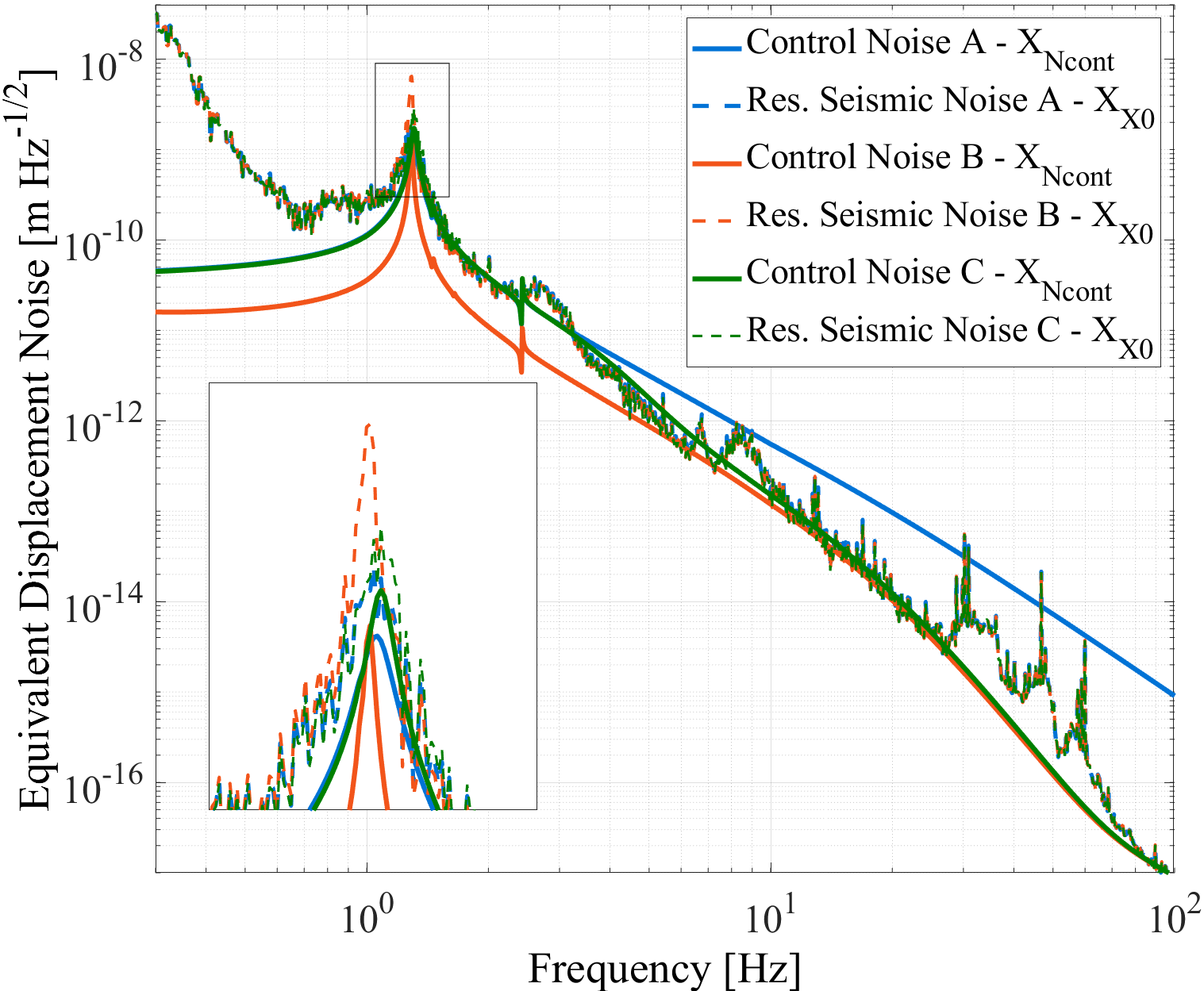}
}

  \caption{\label{fig:HF_cont} \textbf{HIGH FREQUENCY TUNING.} Comparison of three controllers: A, B, and C. (a) Controllers frequency response. Controller A is a band-pass filter. Controller B has high-frequency roll-off and reduced gain. Controller C has a more aggressive roll-off Controller C and has an impact in the phase margin of $\approx$20$^\circ$ (dashed black line corresponds to unity gain frequency for controllers A and C). (b) Noise budget. Estimated motion due to control noise and to input motion for the three controllers. Controller A decreases the platform isolation at 100 Hz by $\approx$40 dB, compared to the undamped situation. Controllers B and C have minimal passive isolation degradation, but the first one presents an increase in motion around the frequency due to the lower gain in that frequency band.}
\end{figure}

Fig. \ref{fig:HF_cont} compares the performance of three example controllers. Controller A is the band-pass filter that has been previously presented. Controllers B and C add high frequency roll-off to prevent high frequency passive isolation deterioration ($X_{N_\mathrm{CONT}} < X_{X_0}$). Controller B has a lower controller gain, which results in a decrease of the damping at the resonance (Fig. \ref{fig:controller_comparison_noise}).  C provides the same level of damping than Controller A but has a more aggressive roll-off after the resonance (Fig. \ref{fig:controller_comparison}). However, this controller is less stable since the phase margin gets reduced by $\approx$20$^\circ$ with respect to Controller A.

In this view, depending on the requirements of the system and the input noise, a compromise must be found between the stability of the controller and its performance at high and low frequency. The development of the controllers used in operations it is made in accordance with the motion requirements for the optical instrument. \cite{chua2013impact}. 
\section{Conclusion}
We presented a compact isolation platform designed to support LIGO auxiliary optics, emphasizing the system's compactness, simplicity, and UHV characteristics. This paper describes the active damping implemented to preserve passive isolation. Experimental measurements show good agreement between predicted and measured performance.  Thanks to its adaptable design and low production cost, this isolation system was easily procured and deployed at the LIGO sites and can be easily adapted to other applications.

\section{Acknowledgments}

We wish to thank Norna Robertson and Calum Torrie
from LIGO-Caltech for their help in designing the suspension, as well as the people from the aLIGO Livinston and Hanford observatories for their help during testing and installation.

LIGO was constructed by the California Institute of Technology and Massachusetts Institute of Technology with funding from the National Science Foundation, and operates under cooperative agreement PHY0757058. Advanced LIGO was built under award PHY0823459. This paper was assigned LIGO Document Number LIGO-P1800182.


\bibliography{bibliog}

\end{document}